\documentclass[12pt]{article}
\usepackage{a4wide}
\usepackage{amssymb}
\usepackage{amsmath}
\usepackage{graphicx}
\usepackage{slashed}
\usepackage{verbatim}

\def\XXint#1#2#3{{\setbox0=\hbox{$#1{#2#3}{\int}$}
     \vcenter{\hbox{$#2#3$}}\kern-.52\wd0}}

\newcommand{\be}{\begin{equation}}\newcommand{\ee}{\end{equation}}
\newcommand{\bea}{\begin{eqnarray}} \newcommand{\eea}{\end{eqnarray}}

\def\makeatletter{\catcode`\@=11}% 11:letter
\makeatletter
\def\mathbox#1{\hbox{$\m@th#1$}}%
\def\math@ccstyles#1#2#3#4#5#6#7{{\leavevmode
      \setbox0\mathbox{#6#7}%
      \setbox2\mathbox{#4#5}%
      \dimen@ #3%
      \baselineskip\z@\lineskiplimit#1\lineskip\z@
      \vbox{\ialign{##\crcr
             \hfil \kern #2\box2 \hfil\crcr
             \noalign{\kern\dimen@}%
             \hfil\box0\hfil\crcr}}}}
\def\mathaccstyles{\math@ccstyles\maxdimen}
\def\maththroughstyles{\math@ccstyles{-\maxdimen}}
\def\unity%
 {\maththroughstyles{.45\ht0}\z@\displaystyle {\mathchar"006C}\displaystyle 1}

\makeatletter \@addtoreset{equation}{section} \makeatother

\begin{document}

\setcounter{table}{0}

\begin{flushright}\footnotesize

\texttt{ICCUB-17-015}
\vspace{0.6cm}
\end{flushright}

\mbox{}
\vspace{0truecm}
\linespread{1.1}

%%%%%%%%%%%%%%%%%
\centerline{\LARGE \bf Free energy and boundary  anomalies}
\medskip

\centerline{\LARGE \bf on $\mathbb{S}^a\times \mathbb{H}^b$ spaces}

\vspace{.5cm}

 \centerline{\LARGE \bf }

\vspace{1.5truecm}

\centerline{
    {\large \bf Diego Rodriguez-Gomez${}^{a}$} \footnote{d.rodriguez.gomez@uniovi.es}
    {\bf and}
    {\large \bf Jorge G. Russo ${}^{b,c}$} \footnote{jorge.russo@icrea.cat}}

\vspace{1cm}
\centerline{{\it ${}^a$ Department of Physics, Universidad de Oviedo}} \centerline{{\it Avda.~Calvo Sotelo 18, 33007  Oviedo, Spain}}
\medskip
\centerline{{\it ${}^b$ Instituci\'o Catalana de Recerca i Estudis Avan\c{c}ats (ICREA)}} \centerline{{\it Pg.Lluis Companys, 23, 08010 Barcelona, Spain}}
\medskip
\centerline{{\it ${}^c$ Departament de F\' \i sica Cu\' antica i Astrof\'\i sica and Institut de Ci\`encies del Cosmos}} \centerline{{\it Universitat de Barcelona, Mart\'i Franqu\`es, 1, 08028
Barcelona, Spain }}
\vspace{1cm}

\centerline{\bf ABSTRACT}
\medskip
We compute free energies as well as conformal anomalies associated with boundaries for a conformal free scalar field. To that matter, we introduce the family of spaces of the form $\mathbb{S}^a\times \mathbb{H}^b$, which are conformally related to $\mathbb{S}^{a+b}$. For the case of $a=1$, related to the entanglement entropy across $\mathbb{S}^{b-1}$, we provide some new explicit computations of entanglement entropies at weak coupling. We then 
compute the free energy for spaces $\mathbb{S}^a\times \mathbb{H}^b$ for different values of $a$ and $b$. For spaces $\mathbb{S}^{2n+1}\times \mathbb{H}^{2k}$
we find an exact match with the free energy on $\mathbb{S}^{2n+2k+1}$.
For $\mathbb{H}^{2k+1}$ and $\mathbb{S}^{3}\times \mathbb{H}^{3}$ we find  conformal anomalies originating
from boundary terms.
 We also compute the free energy for strongly coupled theories through holography,
 obtaining similar results.

\noindent

\newpage

\tableofcontents

\section{Introduction}

The response of a conformal field theory (CFT) to curvature encodes important information about the CFT. On general grounds, upon a Weyl rescaling, a CFT on an even-dimensional curved space suffers from anomalies, which are proportional to the central charges which characterize the CFT
\cite{Duff:1977ay,Christensen:1978md}. In turn, in odd dimensional manifolds without boundary, it is well known that there are no  conformal anomalies
--in particular, no invariant term of odd dimension can be constructed. However,  in the presence of boundaries, CFT's may have conformal anomalies originating from boundary terms, both in even and odd dimensional spaces. 
While there have been some early discussions on this and related issues (see  \textit{e.g.}   \cite{Melmed:1988hm,Moss:1988yf,Dowker:1989ue,McAvity:1992fq,Branson:1995cm,Graham:1999pm}), the topic has been only recently revived starting with  \cite{Fursaev:2013mxa,Jensen:2013lxa,Huang:2014pfa,Jensen:2015swa,Herzog:2015ioa,Fursaev:2015wpa,Solodukhin:2015eca} (see also \cite{Fursaev:2016inw,Huang:2016rol,Herzog:2017xha,Takayanagi:2011zk,Miao:2017gyt,Chu:2017aab,Astaneh:2017ghi,Miao:2017aba}).

In this paper we will be interested on explicit calculations of such effects. We will investigate these boundary anomalies for an interesting class of geometries, namely, euclidean spaces of the form $\mathbb{S}^a\times \mathbb{H}^b$, where $\mathbb{S}^a$ is the $a$-dimensional sphere and $\mathbb{H}^b$ is the $b$-dimensional hyperbolic space.\footnote{ $\mathbb{H}^b$  can be viewed as euclidean $AdS_b$. The 
boundary is $\mathbb{S}^{b-1}$.} When $\mathbb{S}^a$ and    
$\mathbb{H}^b$ have the same radii, these spaces are conformally related to $\mathbb{S}^{a+b}$. To see this, we write the metric of the $\mathbb{S}^{a+b}$ as\footnote{In the following, unless otherwise stated, we will set the overall radius $R$ of our geometries to one.}

\begin{equation}
ds^2_{\mathbb{S}^{a+b}}=d\phi^2+ \cos^2\phi\, ds_{\mathbb{S}^a}^2+\sin^2\phi\,ds_{\mathbb{S}^{b-1}}^2\ .
\end{equation}
Introducing  a new coordinate $y$ defined by

\begin{equation}
\tan\phi=\sinh y\ ,
\end{equation}
we have that

\begin{equation}
\label{metriga}
ds^2_{\mathbb{S}^{a+b}}=\frac{1}{\cosh^2y}\Big( ds_{\mathbb{S}^a}^2+dy^2+\sinh^2y\,ds^2_{\mathbb{S}^{b-1}}\Big)\ .
\end{equation}
Then, upon stripping off the conformal factor, in the parenthesis we recognize the metric of $\mathbb{S}^a\times \mathbb{H}^b$. 

Note that the case $a=1$ corresponds to $\mathbb{S}^1\times \mathbb{H}^b$. This space can be conformally mapped to $\mathbb{R}^{1,b}$, covering the causal development of the $\mathbb{S}^{b-1}$ inside it. Using this fact one can argue \cite{Casini:2011kv} that the entanglement entropy across the sphere (which equals minus the sphere free energy in odd dimensions) maps to the thermal entropy in $\mathbb{S}^1\times \mathbb{H}^b$, thus making these spaces particularly relevant.  The family $\mathbb{S}^a\times \mathbb{H}^b$ is then a natural generalization of $\mathbb{S}^1\times \mathbb{H}^b$. Note that, being conformal to the sphere $\mathbb{S}^{a+b}$, this family can also be conformally mapped to $\mathbb{R}^{a+b}$.

Another interesting case is $b=2$, as it can be regarded as the near-horizon geometry of an extremal black hole. In this case one identifies the $\mathbb{S}^1$ inside $\mathbb{H}^2$ as the thermal circle. Then, by  a thermodynamic argument one  identifies the  free energy on $\mathbb{S}^{2k+2}$ with the corresponding thermal entropy  on $\mathbb{S}^{2k}\times \mathbb{H}^2$ \cite{Mann:1997hm,Solodukhin:2010pk}.

For our purposes, let us differentiate bulk anomalies from boundary anomalies. Bulk anomalies appear only in even dimensions and
are represented by conformal invariant terms constructed in terms of the Riemann tensor and covariant derivatives. This includes the $a$-anomaly, given by
the Euler density, and $c$-anomalies, constructed in terms of the Weyl tensor.
In the presence of boundaries, there are many other conformal invariant terms
containing the extrinsic curvature that can contribute to the integrated trace of the stress tensor \cite{Dowker:1989ue,Fursaev:2013mxa,Herzog:2015ioa,Solodukhin:2015eca}.

Since for the class of spaces $\mathbb{S}^a\times \mathbb{H}^b$ the Weyl tensor
vanishes, bulk conformal anomalies can only appear when the Euler number
of the space is non-vanishing, namely when both $a$ and $b$ are even numbers.
Denoting the free energy for a CFT on $\mathbb{S}^a\times \mathbb{H}^b$ as $F_{(a,b)}$ (note that $(a+b,0)$ corresponds to the $\mathbb{S}^{a+b}$ case), in the absence of conformal anomalies one would have $F_{(a,b)}=F_{(a+b,0)}$ for all $a,b$.\footnote{We should be more precise here, as the free energy suffers from  ambiguities coming from scheme-dependent, regulated UV and IR divergences. As discussed below, here we focus on the logarithmic term and, when this is absent, on the finite part of $F$.}  A mismatch $F_{(a,b)}\neq F_{(a+b,0)}$ for some $a$ and $b$ can only occur when conformal anomalies are present.
In particular, when  $a+b$ is an  odd number 
or else if $a$ is an odd number, 
the CFT can only have 
boundary conformal anomalies.\footnote{There can be also $D$-type anomalies. However, these can be absorbed into local counterterms in the action. See \textit{e.g} \cite{Huang:2013lhw,Beccaria:2014xda} and references therein for discussions.} 

In this paper we will compute  $F_{(a,b)}$ for a number of cases. The calculation will be done  at zero coupling by considering free conformal scalars on $\mathbb{S}^a\times \mathbb{H}^b$.
We also discuss the holographic (strong coupling) calculation   for theories admitting a holographic dual in terms of Einstein gravity. 

Regularizing with a UV cut-off, in the presence of bulk conformal anomalies 
the free energy will be proportional to $\log R\Lambda$. Since $R$ is the overall scale of the geometry, a constant Weyl re-scaling induces a re-scaling of $R$. Thus, the free energy is sensitive to the conformal anomaly.
Specifically, the counterterm that eliminates the logarithmic divergence is the one responsible for the trace anomaly in the stress tensor and the coefficient in front of the $\log $ term is directly related to the anomaly.

The structure of this paper is as follows: in section \ref{preliminaries} we set up the computation of the free energy on $\mathbb{S}^a\times \mathbb{H}^b$ for free scalars. Since this will be compared against the $\mathbb{S}^{a+b}$ free energy, in the same section we  review the  results for $F_{(a+b,0)}$. In section \ref{weakcoupling} we compute the free energies $F_{(a,b)}$ for a number of cases. We first start with $a=1$ and explicitly perform the computation of $F_{(1,b)}$ for $b=2,\cdots,7$. While the cases of $b=2,3$ have appeared in \cite{Klebanov:2011uf}, the cases $b=4,\cdots,7$ offer new explicit and remarkable tests of the identity between thermal entropy on $\mathbb{S}^1\times \mathbb{H}^b$ and entanglement entropy --or sphere free energy-- (see also \cite{Beccaria:2017dmw}). We then focus on the cases for which the total dimension $a+b$ is odd or in cases involving odd spheres, where no bulk Weyl anomalies are expected. In these cases we find that $F_{(a,b)}=F_{(a+b,0)}$ as long as $b$ is even. 
In section \ref{strongcoupling} we compute holographically the free energy for a (strongly coupled) CFT admitting a holographic dual in terms of Einstein gravity. 
We finish in section \ref{discussion} with some concluding remarks. Appendix \ref{appVolHb}  reviews the computation of the regularized volume of the hyperbolic space. For completeness,  appendix \ref{appHoloFSphere} provides a review of the holographic computation of the free energy for a CFT on $\mathbb{S}^{a+b}$. Finally, in appendix \ref{appSUSY} we include a proof that the family of spaces $\mathbb{S}^a\times AdS_b$ admits Killing spinors. Hence, supersymmetric theories can be defined on these spaces.

%%%%%%%%%%%%%%%%%%%%%%%
\section{The weak coupling computation: preliminaries}\label{preliminaries}

The action for a conformally coupled scalar on $\mathbb{S}^a\times \mathbb{H}^b$ is

\begin{equation}
S=\int d^d x \sqrt{g}\,\Big(\partial_\mu\phi\partial^\mu\phi+\frac{(d-2)}{4\,(d-1)}{\cal R}\phi^2\Big)\ ,
\end{equation}
where $d=a+b$. Then, the free energy is given by

\begin{equation}
F_{(a,b)}=-\log Z= \frac{1}{2}{\rm Tr}\,\log\,\Big(-\Delta+\frac{(d-2)}{4\,(d-1)}{\cal R}\Big)\ ,
\end{equation}
with  $\Delta$ being the scalar Laplace operator.
For our spaces this becomes
\begin{equation}
\label{fregy}
F_{(a,b)}=\frac{1}{2}\,{\rm Tr}\log (-\Delta+M^2)\ ,
\end{equation}
\begin{equation}
\label{masas}
M^2\equiv \frac{(a-b)(a+b-2)}{4}\ .
\end{equation}
This can be computed as usual by the heat kernel method, or by a more 
explicit sum over eigenvalues of the Laplacian. 

Let us begin by reviewing the case of $\mathbb{S}^{d}$, which plays a  pivotal role in our discussion.
 It is important to recall that the free energy suffers from  UV divergences. Introducing a UV cut-off $\Lambda$, in general $F$ includes a sum over terms proportional to $(R\Lambda)^{d-2n}$ for $n=0,1,\cdots $, plus some finite coefficient. In addition, in even dimensions, there is a logarithmic term. Thus

\begin{equation}
\begin{array}{l c l}
d=2k\,: & & F = \mathcal{A}_d(R\Lambda)^d+\mathcal{A}_{d-2} (R\Lambda)^{d-2} + \cdots + \mathcal{A}_2(R\Lambda)^2 + \mathcal{A}\,\log(R\Lambda)+\mathcal{F}\, , \\
d=2k+1\,: & & F = \mathcal{B}_d(R\Lambda)^d+\mathcal{B}_{d-2} (R\Lambda)^{d-2} + \cdots +\mathcal{B}_1(R\Lambda)^1 +\mathcal{F} \, .
\end{array}
\end{equation}
Counterterms can be added to remove UV divergences.
In even dimensions,  
there is a  logarithmic term, whose coefficient is free of ambiguities. 
In odd dimensions, the finite term depends on the scheme (in particular,
it would change under a shift $\Lambda\to \Lambda+$constant). However, the finite part in odd dimensions is universal
in the sense that it does not change under re-scaling of $\Lambda$ by a constant.

On the other hand, the free energy on a non-compact space such as $\mathbb{S}^a\times \mathbb{H}^b$ suffers, in addition, from IR divergences. Similarly,
there are universal terms once the power-law divergences have been subtracted.
The dimensional regularization of the volume of $\mathbb{H}^b$  used in appendix A automatically leaves the universal terms (a finite term for even $b$, a logarithmic divergent term for odd $b$).
In odd dimensional spaces of the form  $\mathbb{S}^{2n+1}\times \mathbb{H}^{2k}$ we will be able to match the finite part of the free energy with the finite part in 
$\mathbb{S}^{2n+2k+1}$. This of course requires using the same scheme for subtracting power divergences on both sides.

A particular case in this family are the spaces $\mathbb{S}^{1}\times \mathbb{H}^{2k+1}$, which have been extensively studied in the literature, starting with \cite{Casini:2011kv}. As we will review below, the free energy has no
UV logarithmic divergences in this space. Instead, there is an IR logarithmic divergence
originating from the volume of $\mathbb{H}^{2k+1}$. In connecting the results
on $\mathbb{S}^{1}\times \mathbb{H}^{2k+1}$ with the results on  $\mathbb{S}^{2k+2}$
one uses an UV/IR connection of the cutoffs, implied by the conformal map between both spaces \cite{Casini:2011kv} (see  also \cite{Klebanov:2011uf,Herzog:2015ioa}).
One can extend this idea to the more general family of spaces $\mathbb{S}^{2n+1}\times \mathbb{H}^{2k+1}$, or to spaces $\mathbb{H}^{2k+1}$. There are no UV logarithmic divergences on these spaces, but, likewise, there is an IR logarithmic divergence
originating from the volume of $\mathbb{H}^{2k+1}$.
In comparing with the results of the sphere, one may similarly expect
that there is a relation between the corresponding IR and UV cutoffs implied by the conformal map. We give an argument on appendix \ref{appVolHb}.

The free energy $F_{(d,0)}=-\log Z$ for a conformal free scalar on $\mathbb{S}^d$ is given by \cite{Giombi:2014xxa}

\begin{equation}
F_{(d,0)}=-\frac{1}{\sin\big(\frac{\pi d}{2}\big)\,\Gamma(1+d)}\int_0^1du\,u\,\sin(\pi u)\,\Gamma\big(\frac{d}{2}+u\big)\,\Gamma\big(\frac{d}{2}-u\big)\ .
\end{equation}
In this paper we will explicitly study the cases $d=3,...,8$. The free energy on the sphere $\mathbb{S}^{a+b}$ is

\begin{equation}
\label{segunda}
\begin{array}{l c l}
F_{(3,0)}= \frac{\log 2}{8}-\frac{3}{16\pi^2}\zeta(3)\ , & &F_{(4,0)}= \frac{1}{90\,\epsilon}\ ,\\
\\
F_{(5,0)}= -\frac{\log 2}{128}-\frac{1}{128\pi^2}\zeta(3)+\frac{15}{256\pi^4}\zeta(5)\ , && F_{(6,0)}= -\frac{1}{756\,\epsilon}\ , \\
\\
F_{(7,0)} = \frac{\log 2}{1024}+\frac{41}{30720\pi^2}\zeta(3)-\frac{5}{2048\pi^4}\zeta(5)-\frac{63}{4096\pi^6}\zeta(7) \, , && F_{(8,0)} =\frac{23}{113400\,\epsilon}\ .
\end{array}
\end{equation}
The even case was obtained by doing dimensional regularization $d\rightarrow d-\epsilon$ in the $\epsilon\rightarrow 0^+$ limit, where the $\log (R\Lambda )$ gets traded by the $\frac{1}{\epsilon}$ pole. The even case reflects the $a$ conformal anomaly, which is proportional to the Euler characteristic of $\mathbb{S}^d$ ($\chi(\mathbb{S}^d)=2$ for  $d$ even, $\chi(\mathbb{S}^d)=0$ for $d$ odd).

\subsection{The heat kernel}

The free energy can be computed from the formula:

\begin{equation}
\label{calor}
F_{(a,b)}=-\frac{1}{2}\int_{\delta}^{\infty}\frac{dt}{t}\,K_{\mathbb{H}^b}\,K_{\mathbb{S}^a}\,e^{-M^2t}\ ,
\end{equation}
where $\delta$ is a UV regulator, 
$K_{\mathbb{H}^b}$, $K_{\mathbb{S}^a}$ are the heat kernels for Laplace operators of scalar fields on  $\mathbb{H}^b$ and $\mathbb{S}^a$ spaces, respectively.

 Heat kernels in spheres and hyperbolic spaces have been extensively discussed in the literature. 
 Formulas for the heat kernel on  hyperbolic spaces for real scalars are given \textit{e.g.} in appendix A of \cite{Belin:2013uta}. For odd $b=2n+1$, the general formula reads

\begin{equation}
K_{\mathbb{H}^{2n+1}}=e^{-n^2 t}\,\frac{1}{(4\pi t)^{\frac{1}{2}}}\,\Big(\frac{-1}{2\pi \sinh\rho}\partial_{\rho}\Big)^n e^{-\frac{\rho^2}{4t}}\ .
\end{equation}
The equal-point kernel is obtained
by setting $\rho=0$ and integrating over the $\mathbb{H}^b$ space. 
For future reference, let us quote the result for $b=3,5,7$:

\begin{equation}
K_{\mathbb{H}^3} = \frac{V_{\mathbb{H}^3}}{(4\pi t)^{\frac{3}{2}}}e^{- t}\ ,\quad
K_{\mathbb{H}^5} = \frac{V_{\mathbb{H}_5}}{(4\pi t)^{\frac{5}{2}}}\, (1+\frac{2t}{3})\,e^{-4t}\ ,\quad
K_{\mathbb{H}^7} = \frac{V_{\mathbb{H}_7}}{(4\pi t)^{\frac{7}{2}}} (1+2t+\frac{16}{15}t^2)\,e^{-9t}\ .
\label{heathyp}
\end{equation}
The even $b$ case can also be found in appendix A of \cite{Belin:2013uta}. However, for $b$ even, as the heat kernel is slightly more complicated, it is 
more convenient to compute the determinant directly
from the known expressions of the eigenvalues of the Laplace operator, as explained below.

Similar formulas for the heat kernel on spheres
can be found, \textit{e.g.}, in section (2.1) of \cite{Lewkowycz:2012qr}. Here we quote the cases that will be used in this paper:

\begin{eqnarray}
&& K_{\mathbb{S}^1}=\frac{\beta}{(4\pi t)^{\frac{1}{2}}}\sum_{n\ne 0}e^{-\frac{n^2\beta^2}{4t}}\, ,\qquad
K_{\mathbb{S}^2} = \frac{V_{\mathbb{S}^2}}{4\pi}\sum_{n=0}^\infty (2n+1)\,e^{-n(n+1)t}\ ,\nonumber\\
&& K_{\mathbb{S}^3}= \frac{V_{\mathbb{S}^3}}{2\pi^2}\sum_{n=0}^\infty (n+1)^2 e^{-n(n+2)t}\ .
\end{eqnarray}
The formula for $ K_{\mathbb{S}^1}$ is obtained after Poisson resummation.
For the $\mathbb{S}^3$ case, we used the simpler form given in    \cite{David:2009xg}, 
which uses the fact that the $\mathbb{S}^3$ is the $SU(2)$ group manifold. Note also that the heat kernel for the $\mathbb{S}^1$ assumes a generic length $\beta$ for the circle.

%%%%%%%%%%%%%%%%%%%%%%%%%%%%%%%%%%%%
\subsection{Summing over eigenvalues}

Alternatively, the determinants can also be
computed  from the explicit expression of the eigenvalues of the Laplace operator and their degeneracy.
We first write

\begin{equation}
\Delta=\Delta_{\mathbb{S}^a}+\Delta_{\mathbb{H}^b}\ .
\end{equation}
For scalar fields, the eigenvalues of the Laplacian on $\mathbb{S}^a$ are $l(l+a-1)$ ($l=0,1,\cdots$), with degeneracy $\mathbf{d}^{(a)}_l$ given by

\begin{equation}
\mathbf{d}^{(a)}_{l}=(2l+a-1)\,\frac{\Gamma(l+a-1)}{\Gamma(a)\,\Gamma(l+1)}\ .
\end{equation}

For $\mathbb{H}^b$, we consider the metric (cf. (\ref{metriga}))

\begin{equation}
ds^2_{\mathbb{H}^b}=dy^2+\sinh^2y\ ds^2_{\mathbb{S}^{b-1}}\, ,
\end{equation}
where the boundary is the sphere $\mathbb{S}^{b-1}$.
The  Laplace operator and the density of states on this space $\mathbb{H}^b$ 
have been studied in \cite{Camporesi:1994ga,Bytsenko:1994bc}.
The eigenvalues will be denoted by the continuous  variable $\lambda$, with a density of eigenvalues $\Phi_{(b)}$ given by 

\begin{equation}
\label{degi}
\Phi_{(b)}\,d\lambda=\frac{V_{\mathbb{H}^b}}{(4\pi)^{\frac{b}{2}}\,\Gamma(\frac{b}{2})}\,\frac{|\Gamma(i\,r+\rho_b)|^2}{|\Gamma(i\,r)|^2} \frac{d\lambda}{\sqrt{\lambda-\rho_b^2}}\ ,
\end{equation}
with
$$
r\equiv \sqrt{\lambda-\rho_b^2}\ , \qquad \rho_b\equiv \frac{b-1}{2}\ .
$$
For future reference, let us quote the explicit forms of the densities which we will need

\begin{equation}
\begin{array}{l c l}
\Phi_{(2)} =\frac{V_{\mathbb{H}^2}}{4\pi}\, \tanh\Big(\pi\sqrt{\lambda-\frac{1}{4}}\Big)\, , & & \Phi_{(3)} =\frac{V_{\mathbb{H}^3}\, \sqrt{\lambda-1}}{4\pi^2}\  , \\
\Phi_{(4)}  =\frac{V_{\mathbb{H}^4}}{16\pi^2}\, (\lambda-2)\,\tanh\big(\pi\sqrt{\lambda-\frac{9}{4}}\big)\ , &&  \Phi_{(5)} =\frac{V_{\mathbb{H}^5}\, (\lambda-3)\,\sqrt{\lambda-4}}{24\pi^3}\  ,  \\ 
\Phi_{(6)} =\frac{V_{\mathbb{H}^6}}{128\pi^3}\,(24-10\lambda+\lambda^2)\,\tanh\left( \pi\sqrt{\lambda-\frac{25}{4}}\right), &&  \Phi_{(7)} =\frac{V_{\mathbb{H}^7}\, (40-13\lambda+\lambda^2)\,\sqrt{\lambda-9}}{240\pi^4}\ \,.
\end{array}
\label{eigendens}
\end{equation}
Substituting these expressions into  (\ref{fregy}) and (\ref{masas}), we thus obtain

\begin{equation}
\label{freegen}
F_{(a,b)}=\frac{1}{2}\sum_{l=0}^{\infty} \mathbf{d}^{(a)}_l\,\int_{\rho_b^2}^{\infty} d\lambda\  \Phi_{(b)}\,\log\Big(\lambda+\frac{(a-b)(a+b-2)}{4}+l(l+a-1)\Big)\ .
\end{equation}
The free energy will still depend on the volume of the $\mathbb{H}^b$ space. This volume is divergent but it can be regularized \cite{Diaz:2007an} as discussed in appendix \ref{appVolHb}.

%%%%%%%%%%%%%%%%%%%%%%%%%%%%%%%%%%%%%%%
\section{Free conformal scalar on $\mathbb{S}^a\times \mathbb{H}^b$}\label{weakcoupling}

%%%%%%%%%%%%%%%%%%%%%%%%%%%%%%%%%
\subsection{ Spaces of the form $\mathbb{S}^1\times \mathbb{H}^b$}

Let us begin by considering free real scalar fields on $\mathbb{S}^1\times \mathbb{H}^b$. 
In this case we can view the $\mathbb{S}^1$ as a thermal circle.\footnote{It has been proposed \cite{Zhou:2015kaj} that the Renyi entropy is the same
computed on geometries $\mathbb{S}^n_q \times \mathbb{H}^{d-n}$, for any $n$, where $\mathbb{S}^n_q$ is a branched sphere obtained by a conic defect in any circle of  $\mathbb{S}^n$. 
In the presence of conformal anomalies, it is very unclear whether this could hold  in general, aside from the known $\mathbb{S}^1_q \times \mathbb{H}^{d-1}$  case, but  it would be interesting to explore this proposal by explicit calculations.}
%{We do not expect this to hold
%in general, except for the $n=1$ cases computed here, but it would %be extremely interesting to explore this proposal by explicit %calculations.  }
 Assuming its length to be $\beta$ (note that only the case $\beta=2\pi$ is conformally related to $\mathbb{S}^{1+b}$), one may compute a $\beta$-dependent free energy. Then, it follows that the associated thermal entropy coincides with the entanglement entropy across a spherical surface in $\mathbb{R}^{1,b-2}$  \cite{Casini:2011kv}, 

\begin{equation}
\label{entang}
S_{(1,b)}=-F_{(1,b)}+\beta\partial_{\beta}F_{(1,b)}\Big|_{\beta=2\pi}\ .
\end{equation}
This entanglement entropy is, in turn, equal to minus the $\mathbb{S}^{1+b}$ partition function as $S_{(1,b)}=-F_{(1+b,0)}$. We may then combine these ingredients to write

\begin{equation}
\label{relF}
F_{(1+b,0)}=F_{(1,b)}-\beta\partial_{\beta}F_{(1,b)}\Big|_{\beta=2\pi}\ .
\end{equation}
Typically, for spaces of odd dimension $\beta\partial_{\beta}F_{(1,b)}\Big|_{\beta=2\pi}$ identically vanishes \cite{Klebanov:2011uf} (see also \cite{Dowker:2010yj}).\footnote{It does not always vanish in higher spin examples \cite{Beccaria:2017lcz}.}

For these spaces, the mass coming from the conformal coupling to curvature is

\begin{equation}
M^2=-\frac{(b-1)^2}{4}\ .
\end{equation}
By shifting the integration variable $\lambda-  \frac{(b-1)^2}{4}\to \lambda $,
the free energy (\ref{freegen}) takes the form
\begin{equation}
\label{freegen2}
F_{(1,b)}=\frac{1}{2}\sum_{l=-\infty}^{\infty} \,\int_{0}^{\infty}d\lambda\  \Phi_{(b)}\,\log\Big(\lambda + l^2\Big)\ .
\end{equation}
The sum over $l$ can be computed using the product representation of $\sinh$ (see \textit{e.g.} appendix B in \cite{Klebanov:2011uf}). 
We obtain 
\begin{equation}
\label{freegen3}
F_{(1,b)}= \,\int_{0}^{\infty}d\lambda \ \Phi_{(b)}\,\log\Big(2\sinh(\pi \sqrt{\lambda})\Big)\ .
\end{equation}
The formula is readily generalized to the case of an $\mathbb{S}^1$ 
of length $\beta = 2\pi q $ by replacing 
$\sinh(\pi \sqrt{\lambda})\to \sinh(\pi q  \sqrt{\lambda})$.

In the $b$ odd case, the free energy can be most directly computed by using the heat kernel.
Note that when $b=2k+1$, one has $M^2=-k^2$ and the conformal coupling to curvature exactly cancels the $t$ exponential in the $\mathbb{H}^b$ heat kernel.

\subsubsection{$\mathbb{S}^1\times \mathbb{H}^2$}

We first compute the free energy for $\beta =2\pi$, following \cite{Klebanov:2011uf}. Making use of the explicit expression \eqref{eigendens} for the eigenvalue density, the free energy \eqref{freegen3} takes the form

\begin{equation}
F_{(1,2)}=\frac{V_{\mathbb{H}^2}\, }{4\pi} \int_{0}^{\infty}d\lambda\  \tanh\big(\pi\sqrt{\lambda}\big)\,\log\big(2\sinh(\pi\sqrt{\lambda})\big)\ .
\end{equation}
This can be written in the form:

\begin{equation}
F_{(1,2)}=\frac{V_{\mathbb{H}^2}\, }{4\pi} \int_{0}^{\infty} d\lambda\  \tanh\big(\pi\sqrt{\lambda}\big)\,\log\Big(1-e^{-2\pi\sqrt{\lambda}}\Big) + \frac{V_{\mathbb{H}^2}\, }{4} \int_{0}^{\infty}d\lambda \  \tanh\big(\pi\sqrt{\lambda}\big)\,\sqrt{\lambda}\ .
\end{equation}
The second integral diverges. 
As in \cite{Klebanov:2011uf}, we regulate it by subtracting the $\mathbb{R}^3$
free energy density:
\begin{equation}
\frac{V_{\mathbb{H}_2}}{4}\int_0^{\infty}d\lambda\,  \sqrt{\lambda}\ .
\end{equation}
Thus, the regularized free energy is

\begin{equation}
F_{(1,2)}=\frac{V_{\mathbb{H}^2}\, }{4\pi} \int_{0}^{\infty}d\lambda\,  \tanh\big(\pi\sqrt{\lambda}\big)\,\log\Big(1-e^{-2\pi\sqrt{\lambda}}\Big) + \frac{V_{\mathbb{H}^2}}{4} \int_{0}^{\infty}d\lambda\,  \Big[\tanh\big(\pi\sqrt{\lambda}\big)-1\Big]\,\sqrt{\lambda}\ .
\nonumber
\end{equation}

The $\lambda$ integrals can be computed using the formulas 

\begin{eqnarray}
&& \frac{V_{\mathbb{H}^2}}{4} \int_{0}^{\infty}d\lambda\  \Big[\tanh\big(\pi\sqrt{\lambda}\big)-1\Big]\,\sqrt{\lambda}=-\frac{3V_{\mathbb{H}^2}\,\zeta(3) }{16\pi^3}\ ,\\
&& \frac{V_{\mathbb{H}^2}\, }{4\pi} \int_{0}^{\infty}d\lambda\  \tanh\big(\pi\sqrt{\lambda}\big)\,\log\Big(1-e^{-2\pi\sqrt{\lambda}}\Big) =-\frac{V_{\mathbb{H}^2}\, }{16\pi} \Big(\log 2-\frac{9\zeta(3)}{2\pi^2}\Big)\ .
\label{inig}
\end{eqnarray}
One finally finds

\begin{equation}
F_{(1,2)}=-\frac{V_{\mathbb{H}^2}}{16\pi}\log 2+\frac{3V_{\mathbb{H}_2}\zeta(3)}{32\pi^3}\ .
\end{equation}
Using now  $V_{\mathbb{H}^2}=-2\pi$, one gets \cite{Klebanov:2011uf} 

\begin{equation}
F_{(1,2)}=\frac{\log 2}{8}-\frac{3\zeta(3)}{16\pi^2}=F_{(3,0)}\ ;
\end{equation}
thus reproducing the result of the free energy on $\mathbb{S}^3$.

The formulas are readily generalized for 
an $\mathbb{S}^1$ of length $\beta = 2\pi q$ (see also \cite{Klebanov:2011uf}). One finds

\begin{equation}
F_{(1,2)}=\frac{V_{\mathbb{H}^2}\, }{4\pi} \int_{0}^{\infty}d\lambda  \tanh\big(\pi\sqrt{\lambda}\big)\,\log\Big(1-e^{-2\pi q\sqrt{\lambda}}\Big) + \frac{q V_{\mathbb{H}^2}}{4} \int_{0}^{\infty} d\lambda  \Big[\tanh\big(\pi\sqrt{\lambda}\big)-1\Big]\,\sqrt{\lambda}\ .
\end{equation}
It then follows that

\begin{equation}
\frac{\partial F_{(1,2)}}{\partial q}\Big|_{q=1}=0\ .
\end{equation}
Thus, in this case, there is no new contribution from the term $\beta\partial_{\beta}F_{(1,b)}$. The same feature holds for all even $b$  cases.\footnote{The 
analogous statement for  $\mathbb{Z}_q$ orbifolds of the sphere has been argued in general in \cite{Dowker:2010yj}.}

\subsubsection{$\mathbb{S}^1\times \mathbb{H}^3$}

Let us now consider the case $\mathbb{S}^1\times \mathbb{H}^3$
(also considered in the Discussion section of \cite{Klebanov:2011uf}).  From \eqref{calor},  \eqref{heathyp},  we have

\begin{equation}
F_{(1,3)}=-\frac{1}{2}\frac{V_{\mathbb{H}^3}\beta}{(4\pi)^2} \sum_{n\ne 0}\int_{\delta}^{\infty}\frac{dt}{t^3} e^{-\frac{n^2\beta^2}{4t}}\ .
\end{equation}
Computing the  integral and the infinite sum, we obtain

\begin{equation}
\label{F13}
F_{(1,3)}=-\frac{\pi^2\,V_{\mathbb{H}^3}}{90\beta^3}\ .
\end{equation}
We can now use \eqref{relF} to find 

\begin{equation}
-S_{(1,3)}=-\frac{V_{\mathbb{H}^3}}{180\pi}=\frac{1}{90\epsilon}\ ,
\label{hug}
\end{equation}
where we have used that $V_{\mathbb{H}^3}=-\frac{2\pi}{\epsilon}$ (see appendix A). Switching from DREG to cut-off regulated quantities, this coincides with the $\mathbb{S}^4$ free energy $F_{(4,0)}$ (cf. eq. (\ref{segunda})), if one identifies the UV cutoff in  $F_{(4,0)}$
with the IR cutoff in \eqref{hug}.
A justification of this identification --based on mapping the two cutoffs by using the conformal map between the two spaces-- is given in section 2 of \cite{Casini:2011kv}
(see also \cite{Klebanov:2011uf,Herzog:2015ioa,Beccaria:2017dmw} and appendix B).

The same result can be obtained by the alternative method
of summing over eigenvalues.
In this case

\begin{equation}
F_{(1,3)}=\frac{V_{\mathbb{H}^3}}{4\pi^2} \int_{0}^{\infty} d\lambda\  \sqrt{\lambda}\,\log\big(2 \sinh(\pi q\sqrt{\lambda})\big)\ .
\end{equation}
That is,
\begin{equation}
F_{(1,3)}=\frac{V_{\mathbb{H}^3}}{4\pi^2} \int_{0}^{\infty}d\lambda\,  \sqrt{\lambda}\,\log\big(1-e^{-2\pi q \sqrt{\lambda}}\big)+\frac{q V_{\mathbb{H}^3}}{4\pi} \int_{0}^{\infty} d\lambda\, \lambda\ .
\end{equation}
The second term diverges.
Like in the case  of $\mathbb{S}^1\times \mathbb{H}^2$,
this divergence can be regularized by subtracting the flat-theory
free energy density. Computing the remaining integral, we reproduce \eqref{F13}.

\subsubsection{$\mathbb{S}^1\times \mathbb{H}^4$}

{}From \eqref{eigendens} and \eqref{freegen3}, 
we obtain the following expression for the free energy for $\beta=2\pi q$:

\begin{equation}
F_{(1,4)}=\frac{V_{\mathbb{H}^4} }{16\pi^2 }\int_{0}^{\infty} d\lambda\, (\lambda+\frac{1}{4}) \tanh( \pi\sqrt{\lambda})\,\log\Big(2\sinh(\pi q \sqrt{\lambda})\Big)\ .
\end{equation}
Again, we first separate the divergent integral by reorganizing the different terms and then subtract the flat-theory
free energy density. This leads to

\begin{eqnarray}
F_{(1,4)} &=&\frac{V_{\mathbb{H}^4} }{16\pi^2 }\int_{0}^{\infty} d\lambda\  (\lambda+\frac{1}{4}) \tanh( \pi\sqrt{\lambda})\,\log\Big(1-e^{-2\pi q\sqrt{\lambda}}\Big)
\nonumber\\
&+&\frac{q V_{\mathbb{H}^4} }{16\pi}\int_{0}^{\infty} d\lambda (\lambda+\frac{1}{4})\, \sqrt{\lambda} \, \big(\tanh( \pi\sqrt{\lambda})-1\big) \, .
\end{eqnarray}
We note that

\begin{equation}
\frac{\partial F_{(1,4)}}{\partial q}\Big|_{q =1}=0\ .
\end{equation}
Thus there is no contribution from the term $\beta\partial_{\beta}F_{(1,b)}$.
Hence in what follows we set $q =1$.
For the first integral, we use (\ref{inig}) and
\begin{equation}
\int_{0}^{\infty} d\lambda\, \lambda \tanh( \pi\sqrt{\lambda})\,\log\Big(1-e^{-2\pi\sqrt{\lambda}}\Big)=
-\frac{3 \zeta (3)}{16 \pi ^2}+\frac{225 \zeta (5)}{64 \pi ^4}-\frac{\log
   (2)}{32}\ .
\end{equation}
Computing the remaining integral, we find
\begin{equation}
F_{(1,4)}=-\frac{\log 2}{128}-\frac{1}{128\pi^2}\zeta(3)+\frac{15}{256\pi^4}\zeta(5)=F_{(5,0)}\ ;
\end{equation}
thus showing the expected match.

\subsubsection{$\mathbb{S}^1\times \mathbb{H}^5$}

Using the heat kernel formula for the free energy 
\eqref{calor}, with the heat kernel $K_{\mathbb{H}^5}$ given in \eqref{heathyp},  we obtain

\begin{equation}
F=-\frac{1}{2}\frac{V_{\mathbb{H}^5}\beta}{(4\pi)^3} \sum_{n\ne 0}\int_{\delta}^{\infty}\frac{dt}{t^4} e^{-\frac{n^2\beta^2}{4t}}\,\Big(1+\frac{2t}{3}\Big)\ .
\end{equation}
Computing the integral and the infinite sum, we now find

\begin{equation}
\label{F15}
F_{(1,5)}=-\frac{\pi\,V_{\mathbb{H}^5}\,(8\pi^2+7\beta^2)}{3780\,\beta^5}\ .
\end{equation}
The entanglement entropy is then computed from \eqref{entang} and then
 setting $\beta=2\pi$. This gives

\begin{equation}
S_{(1,5)}=\frac{V_{\mathbb{H}^5}}{756\pi^2}\ .
\end{equation}
Using that $V_{\mathbb{H}^5}=\frac{\pi^2}{\epsilon}$, we finally obtain

\begin{equation}
S_{(1,5)}=\frac{1}{756\epsilon}\ .
\end{equation}
Thus, if one identifies IR and UV cutoffs (see discussion above),  $-S_{(1,5)}$ matches the free energy $F_{(6,0)}$ of a scalar in $\mathbb{S}^6$, given in (\ref{segunda}).

\subsubsection{$\mathbb{S}^1\times \mathbb{H}^6$}

In this case, the formulas \eqref{freegen3}, \eqref{eigendens}
give

\begin{equation}
F_{(1,6)}=-\frac{1}{240}\int_0^{\infty}d\lambda\  (\lambda+\frac{1}{4})(\lambda+\frac{9}{4})\,\tanh(\pi\sqrt{\lambda})\,\log\Big(2\sinh(\pi q \sqrt{\lambda})\Big)\ .
\end{equation}
As in previous cases, we rewrite the formula by separating the divergent piece representing
the flat-theory free energy and then subtract this divergence. This leads to
\begin{eqnarray}
F_{(1,6)} &=& -\frac{1}{240}\int_0^{\infty} d\lambda (\lambda+\frac{1}{4})(\lambda+\frac{9}{4})\, \bigg(\tanh(\pi\sqrt{\lambda})\, \log\Big(1-e^{-2\pi q \sqrt{\lambda}}\Big)
\nonumber\\
&+& q \pi\sqrt{\lambda}\ (\tanh(\pi\sqrt{\lambda})-1)\bigg)\ .
\label{unoseis}
\end{eqnarray}
As in all even $b$ cases, one finds  that there is no contribution from the term $\beta\partial_{\beta}F_{(1,b)}\Big|_{\beta =2\pi }$,  which vanishes identically.
The integrals in \eqref{unoseis} can be computing by expanding the log and
resumming the result after integration. Setting $q=1$,
we find 

\begin{equation}
F_{(1,6)}= \frac{\log 2}{1024}+\frac{41}{30720\pi^2}\zeta(3)-\frac{5}{2048\pi^4}\zeta(5)-\frac{63}{4096\pi^6}\zeta(7) \ ,
\end{equation}
which exactly agrees with the  free energy $F_{(7,0)}$ on $\mathbb{S}^7$; see (\ref{segunda}).

\subsubsection{$\mathbb{S}^1\times \mathbb{H}^7$}

In this case, the free energy in the heat kernel representation  \eqref{calor},
with $K_{\mathbb{H}^7}$ given in \eqref{heathyp}, takes the form

\begin{equation}
F_{(1,7)}=\frac{V_{\mathbb{H}^5}\beta}{(4\pi)^4} \sum_{n\ne 0}\int_{\delta}^{\infty}\frac{dt}{t^5} e^{-\frac{n^2\beta^2}{4t}}\,\Big(1+2t+\frac{16t^2}{15}\Big)\ .
\end{equation}
Computing the integral and the infinite sum, we now obtain

\begin{equation}
F_{(1,7)}=-\frac{V_{\mathbb{H}^7}\,(6\pi^4+10\pi^2\beta^2+7\beta^4)}{9450\beta^7}\ .
\end{equation}
Hence, the entanglement entropy \eqref{entang} is given by

\begin{equation}
S_{(1,7)}=\frac{23\,V_{\mathbb{H}^7}}{37800\pi^3}=-\frac{23}{113400\,\epsilon}=-F_{(8,0)}\ .
\end{equation}
where we used $V_{\mathbb{H}^7}=-\frac{\pi^3}{3\epsilon}$. Thus, we see that this exactly matches free energy for a conformal scalar field on $\mathbb{S}^8$, assuming that UV and IR cutoffs can be identified by the arguments of \cite{Casini:2011kv,Herzog:2015ioa}  (see discussion above and appendix B).

%%%%%%%%%%%%%%%%%%%
\subsection{Spaces of the form $\mathbb{S}^{2n+1}\times \mathbb{H}^{2k}$}
%%%%%%%%%%%%%%%%%%%

Since in odd dimensions the bulk anomaly vanishes, a conformal anomaly, if present, could only originate from  boundary contributions. Let us first concentrate  on spaces of the form $\mathbb{S}^{2n+1}\times \mathbb{H}^{2k}$. For these cases we shall find that $F_{(2n+2k+1,0)}=F_{(2n+1,2k)}$. The match is striking, since the result involves a non-trivial combination of Riemann $\zeta$-functions, arising after a long calculation that uses expressions which are 
very different from the expressions used for the spheres $\mathbb{S}^{2n+2k+1}$.

%%%%%%%%%%%%%%%%%%%%%%%%%%%%%%%
\subsubsection{$\mathbb{S}^3\times \mathbb{H}^2$}

Our starting point is \eqref{freegen}. The eigenvalues of  the Laplace operator on $\mathbb{S}^3$ have degeneracy  $\mathbf{d}^{(3)}_l =(l+1)^2$.
Therefore

\begin{equation}
F_{(3,2)}=\frac{1}{2}\sum_{l=0}^{\infty}(l+1)^2\,\int_{\frac{1}{4}}^{\infty} d\lambda \  \Phi_{(2)}\,\log\Big(\lambda-\frac{1}{4}+(l+1)^2\Big)\ .
\end{equation}
By  shifting $\lambda$ by $\frac{1}{4}$ and $l$ by 1, one gets

\begin{equation}
F_{(3,2)}=\frac{V_{\mathbb{H}^2}}{8\pi}\sum_{l=1}^{\infty}l^2\,\int_{0}^{\infty}d\lambda\, \tanh(\pi\sqrt{\lambda})\,\log\Big(\lambda+l^2\Big)\ .
\end{equation}
The sum over $l$ can be  regularized as follows. We consider the auxiliary sum:

\begin{equation}
S_1(m^2)=\sum_{l=1}^\infty l^2\,\log\Big(\lambda+l^2+m^2\Big)\ .
\end{equation}
Our original sum is then obtained as $S_1(m=0)$. By differentiating $S_1$ twice  with respect to $m^2$ and performing the sum over $l$, we get
\begin{equation}
\partial_{m^2}^2S_1=\frac{\pi}{8\,(m^2+\lambda)\sinh^2\big(\pi\sqrt{m^2+\lambda}\big)}\Big(2\pi\,(m^2+\lambda)-\sqrt{m^2+\lambda}\sinh\big(2\pi\sqrt{m^2+\lambda}\big)\Big)\ .
\nonumber
\end{equation}
This can be integrated twice (and then $m^2$ set to zero), leading to the formula

\begin{equation}
\Sigma_1(\lambda)\equiv S_1(0)=-\frac{\pi\lambda^{3/2}}{3}
-\lambda\, \log(1-e^{-2\pi\sqrt{\lambda}})
+\frac{\sqrt{\lambda}}{\pi}\,{\rm Li}_{2}(e^{-2\pi\sqrt{\lambda}})+\frac{1}{2\pi^2}\,{\rm Li}_3(e^{-2\pi\sqrt{\lambda}})\ .
\end{equation}
Making use of this result, we obtain 
 
\begin{equation}
F_{(3,2)}=-\frac{1}{4}\int_0^{\infty}d\lambda\,\Sigma_1(\lambda)\,\tanh(\pi\sqrt{\lambda})\ .
\end{equation}
Here we used that $V_{\mathbb{H}^2}=-2\pi$.
The asymptotic $\lambda\rightarrow \infty$ behavior of $\Sigma_1(\lambda)$ is
\begin{equation}
\Sigma^{\infty}_1=\lim_{\lambda\rightarrow \infty}\Sigma_1(\lambda)=-\frac{\pi}{3}\lambda^{\frac{3}{2}}\ .
\end{equation}
Therefore, the regularized free energy is 

\begin{equation}
F_{(3,2)}=-\frac{1}{4}\int_0^{\infty}d\lambda\, \left(\Sigma_1(\lambda)\,\tanh(\pi\sqrt{\lambda})-\Sigma^{\infty}_1\right)\ .
\end{equation}

To compute the integral one can introduce a new integration variable
$x=e^{-2\pi\sqrt{\lambda}}$  and expand the polylogarithms in powers
of  $x$. 
After some algebra, we find
\begin{equation}
F_{(3,2)}= -\frac{\log 2}{128}-\frac{1}{128\pi^2}\zeta(3)+\frac{15}{256\pi^4}\zeta(5) \ ,
\end{equation}
which thus exactly matches the free energy $F_{(5,0)}$ \eqref{segunda} on the $\mathbb{S}^5$.

%%%%%%%%%%%%%%%%%%%%%%%%%
\subsubsection{$\mathbb{S}^3\times \mathbb{H}^4$}

Using \eqref{freegen} and the expression \eqref{eigendens} for the eigenvalue density in $\mathbb{H}^4$, we are led to the formula:

\begin{equation}
F_{(3,4)}=\frac{V_{\mathbb{H}^4}}{32\pi^2}\sum_{l=0}^{\infty}l^2\,\int_{0}^{\infty} d\lambda\  (\lambda+\frac{1}{4})\,\tanh(\pi\sqrt{\lambda})\log\Big(\lambda+l^2\Big)\ .
\end{equation}
Computing the sum over $l$ as in the previous case,  we get

\begin{equation}
F_{(3,4)}=\frac{1}{24}\int_{0}^{\infty} d\lambda\, (\lambda+\frac{1}{4})\,\tanh(\pi\sqrt{\lambda})\,\Sigma_1(\lambda )\ ,
\end{equation}
where we used $V_{\mathbb{H}^4}=\frac{4\pi^2}{3}$ .
Subtracting the flat-theory free energy, we  find the finite integral:

\begin{equation}
F_{(3,4)}=\frac{1}{24}\int_{0}^{\infty}d\lambda\  (\lambda+\frac{1}{4}) \,\left(\tanh(\pi\sqrt{\lambda})\,\Sigma_1(\lambda )- \Sigma^{\infty}_1\right)\ .
\end{equation}
Computing the integrals, we finally get

\begin{equation}
F_{(3,4)}= \frac{\log 2}{1024}+\frac{41}{30720\pi^2}\zeta(3)-\frac{5}{2048\pi^4}\zeta(5)-\frac{63}{4096\pi^6}\zeta(7)  \ .
\end{equation}
We again find exact match with the free energy $F_{(7,0)}$ on $\mathbb{S}^7$, given in \eqref{segunda}.

%%%%%%%%%%%%%%%%%%%%%%%%%%%%%%%%%%%%
\subsubsection{$\mathbb{S}^5\times \mathbb{H}^2$}

From the general formula \eqref{freegen}, we now get

\begin{equation}
F_{(5,2)}=\frac{1}{2}\sum_{l=0}^{\infty}\frac{(l+1)(l+2)^2(l+3)}{12}\,\int_{\frac{1}{4}}^{\infty} d\lambda\ \Phi_{(2)}\,\log\Big(\lambda-\frac{1}{4}+(l+2)^2\Big)\ .
\end{equation}
Using the expression \eqref{eigendens} for the eigenvalue density,  shifting $\lambda$ by $\frac{1}{4}$ and using that $V_{\mathbb{H}^2}=-2\pi$, we obtain

\begin{equation}
F_{(5,2)}=-\frac{1}{48}\sum_{l=0}^\infty (l+1)(l+2)^2(3+l)\,\int_0^{\infty}d\lambda\, \tanh(\pi\sqrt{\lambda})\,\log\Big(\lambda+(l+2)^2\Big)\ .
\end{equation}
The sum over $l$ is regularized along the same lines as above. Introduce now the auxiliary sum

\begin{equation}
S_2(m^2)=\sum_{l=0}^\infty (l+1)(l+2)^2(3+l)\,\log\Big(\lambda+m^2+(l+2)^2\Big)\ .
\end{equation}
It can be calculated by differentiating three times with respect to $m^2$,
i.e. by considering $\partial_{m^2}^3 S_2(m^2)$, then integrating three times and setting $m^2=0$.
We find

\begin{eqnarray}
\Sigma_2(\lambda)\equiv S_2(0)&=&\frac{\pi \lambda^{\frac{3}{2}}}{3}+\frac{\pi \lambda^{\frac{5}{2}}}{5}-\lambda(1+\lambda)\,{\rm Li}_1(e^{-2\pi\sqrt{\lambda}})-\frac{\lambda^{\frac{1}{2}}(1+2\lambda)}{\pi}\,{\rm Li}_2(e^{-2\pi\sqrt{\lambda}})\nonumber \\  &&  -\frac{(1+6\lambda)}{2\pi^2}\,{\rm Li}_3(e^{-2\pi\sqrt{\lambda}})-\frac{3\lambda^{\frac{1}{2}}}{\pi^3}\,{\rm Li}_4(e^{-2\pi\sqrt{\lambda}})-\frac{3}{2\pi^4}{\rm Li}_5(e^{-2\pi\sqrt{\lambda}})\ .
\end{eqnarray}
Making use of this,  we have

\begin{equation}
F_{(5,2)}=-\frac{1}{48}\int_0^{\infty}d\lambda\,\Sigma_2(\lambda)\,\tanh(\pi\sqrt{\lambda})\ .
\end{equation}
The asymptotic behavior of $\Sigma_2$ for large $\lambda$ is

\begin{equation}
\Sigma^{\infty}_2=\frac{\pi}{3}\lambda^{\frac{3}{2}}+\frac{\pi}{5}\lambda^{\frac{5}{2}}\ .
\end{equation}
Thus, the regularized free energy is given by

\begin{equation}
F_{(5,2)}=-\frac{1}{48}\int_0^{\infty}d\lambda\, \left(\Sigma_2(\lambda)\,\tanh(\pi\sqrt{\lambda})-\Sigma^{\infty}_2\right)\ .
\end{equation}
Computing the integrals, we finally find

\begin{equation}
F_{(5,2)}=\frac{\log 2}{1024}+\frac{41}{30720\pi^2}\zeta(3)-\frac{5}{2048\pi^4}\zeta(5)-\frac{63}{4096\pi^6}\zeta(7) =F_{(7,0)} \ ,
\end{equation}
which, strikingly, exactly matches the free energy $F_{(7,0)}$ \eqref{segunda} on $\mathbb{S}^7$.

%%%%%%%%%%%%%%%%%%%%%%%%%%%%%%%%%%%%%%%%
\subsection{Spaces of the form $\mathbb{S}^{2n}\times \mathbb{H}^{2k+1}$}
%%%%%%%%%%%%%%%%%%%%%%%%%%%%%%%%%%%%%%%%

\subsubsection{$\mathbb{H}^{2k+1}$}

The simplest subclass in  $\mathbb{S}^{2n}\times \mathbb{H}^{2k+1}$  is when $n=0$. The spaces $\mathbb{H}^{2k+1}$ are of course also related to $\mathbb{S}^{2k+1}$
by a Weyl transformation (the precise Weyl transformation is easily found by recalling that both spaces are  conformally flat).
Below it will be shown that, in this case,  $F_{(2k+1,0)}\ne F_{(0,2k+1)}$. While $F_{(2k+1,0)}$, corresponding to the free energy of an odd sphere, is a transcendental number involving zeta functions, see \eqref{segunda}, in contrast $F_{(0,2k+1)}$ contains a $\frac{1}{\epsilon}$ divergence (in DREG --or a $\log \rho_0 $ in terms of the IR cutoff) multiplying a rational number. Since this implies a logarithmic dependence on the scale, it suggests the presence of a boundary conformal anomaly
for the conformal field theory on $\mathbb{H}^{2k+1}$ (below this will be confirmed
independently for  $\mathbb{H}^{3}$).

\paragraph{$\mathbb{H}^3$}:

In this case the conformal mass is $M^2=-\frac{3}{4}$. From \eqref{calor} and \eqref{heathyp}, we find that the free energy on $\mathbb{H}^3$ is
given by

\begin{equation}
F_{(0,3)}=-\frac{1}{2}\frac{V_{\mathbb{H}^3}}{(4\pi)^{\frac{3}{2}}}\int_{\delta}^\infty \frac{dt}{t^{\frac{5}{2}}}e^{-\frac{t}{4}}\ .
\end{equation}
This contains power-law divergences as $\delta\to 0$. The regularized free energy is obtained by subtracting these terms:

\begin{equation}
F_{(0,3)}=-\frac{1}{2}\frac{V_{\mathbb{H}^3}}{(4\pi)^{\frac{3}{2}}}\int_{0}^\infty \frac{dt}{t^{\frac{5}{2}}}\left(e^{-\frac{t}{4}}-1+\frac{t}{4}\right)
= -\frac{V_{\mathbb{H}^3}}{96\pi}\ .
\end{equation}
 Thus, we have 

\begin{equation}
\label{Fcerotres}
F_{(0,3)}=\frac{1}{48}\frac{1}{\epsilon}\ .
\end{equation}
This agrees with the result of \cite{Giombi:2008vd}. 

We can reproduce the same result from the sum over eigenvalues. From \eqref{freegen}, after a shift in the $\lambda$ integration variable, we get

\begin{equation}
F_{(0,3)}=\frac{V_{\mathbb{H}^3}}{8\pi^2}\int_0^{\infty}d\lambda\, \sqrt{\lambda}\, \log(\lambda+\frac{1}{4})\, .
\label{infe}
\end{equation}
We will regulate the divergent integral by zeta function regularization.
We consider
\begin{equation}
\frac{V_{\mathbb{H}^3}}{8\pi^2}\int_0^{\infty}d\lambda\, \sqrt{\lambda}\ \left(\lambda+\frac{1}{4}\right)^{-s}=\frac{2^{2s-7}V_{\mathbb{H}^3}}{\pi^{\frac{3}{2}}} \frac{\Gamma\big(s-\frac{3}{2}\big)}{\Gamma(s)}\, ,\qquad s>\frac{3}{2}\ .
\end{equation}
The integral \eqref{infe} is then obtained by analytic continuation to the
$s\to 0 $ limit, as the coefficient of 
 $-s$ in the expansion in powers of $s$. This reproduces \eqref{Fcerotres}.

We may relate this result to the anomalous trace of the stress energy tensor. A  pole in $\epsilon$ in DREG for the volume of $\mathbb{H}^3$ corresponds to a term 
$\log \rho_0$, $1/\epsilon\to \log \rho_0$, if one regulates the IR divergence in terms of a cut-off $\rho_0 $
(see appendix A). 
The presence of a $\log\rho_0 $ term (given the UV/IR connection discussed in appendix A) suggests the existence of a conformal anomaly due to boundary terms (note that a constant Weyl scaling can be implemented by
scaling $\rho_0$). 
The contribution of boundary terms to the trace of the stress tensor
has been computed in \cite{Solodukhin:2015eca,Fursaev:2016inw}. While these formulas have been derived for compact spaces with boundary, we will naively extend them to our regularized hyperbolic space. 
The general formula for the conformal anomaly containing the boundary contribution is (see section 5 in \cite{Solodukhin:2015eca})

\begin{equation}
\int_{\mathcal{M}_3} \langle T^{\mu}_{\mu}\rangle=\frac{c_1}{96}\chi(\partial\mathcal{M}_3)+\frac{c_2}{256\pi}\int_{\partial\mathcal{M}_3}{\rm Tr}\,\hat{\Theta}\, ,
\label{traso}
\end{equation}
where, for a conformally coupled scalar with Dirichlet boundary conditions, $c_1=-1$, $c_2=1$. Here $\hat{\Theta}$ is the trace-free extrinsic curvature of the boundary metric and $\chi(\partial \mathcal{M}_3)$ is the Euler number of the boundary metric. As for the former, recall that the $\mathbb{H}^3$ metric is $ds^2=dy^2+\sinh^2y\, d\Omega_2^2$. Introducing $\sinh y=\rho $, the metric becomes $ds^2=(1+\rho^2)^{-1}d\rho^2+\rho^2 d\Omega_2^2$. Thus, we can borrow the computation from footnote 10 below with $b=0$, which shows that the trace-free part of $\Theta$ vanishes. Since, on the other hand the boundary is an $\mathbb{S}^2$, for which $\chi(\mathbb{S}^2)=2$, we find 
$$
\int_{\mathbb{H}^3} \langle T^{\mu}_{\mu} \rangle = -\frac{1}{48}\ .
$$
Thus, naive application of  the formulas in \cite{Solodukhin:2015eca,Fursaev:2016inw} shows the CFT on $\mathbb{H}^3$
has a conformal anomaly due to boundary terms, whose coefficient is consistent with \eqref{Fcerotres}.

\paragraph{$\mathbb{H}^5$}:

In this case $M^2=-\frac{15}{4}$. Substituting the heat kernel for  $\mathbb{H}^5$, given in \eqref{heathyp}, into the free energy \eqref{calor}, we obtain

\begin{equation}
F_{(0,5)}=-\frac{1}{2}\frac{V_{\mathbb{H}^5}}{(4\pi)^{\frac{5}{2}}}\int_{\delta}^\infty \frac{dt}{t^{\frac{7}{2}}}(1+\frac{2}{3}t)\, e^{-\frac{t}{4}}\ .
\end{equation}
Regularizing the integral as before to subtract the power-law divergent terms in $\delta $, we find 

\begin{equation}
\label{Fcerocinco}
F_{(0,5)}= -\frac{17}{11520}\frac{1}{\epsilon}\ .
\end{equation}

Likewise, we can recover the same result from explicitly summing over eigenvalues. From \eqref{freegen}, we now get

\begin{equation}
\label{freeci}
F_{(0,5)}=\frac{V_{\mathbb{H}^5}}{48\pi^3}\int_0^{\infty}d\lambda\, \sqrt{\lambda}\ (1+\lambda)\,\log(\lambda+\frac{1}{4})\, .
\end{equation}

The integral can be computed by zeta-function regularization as above. 
We consider
\begin{equation}
\frac{V_{\mathbb{H}^5}}{48\pi^3}\int_0^{\infty}d\lambda\, \sqrt{\lambda}\ (1+\lambda)\,\left(\lambda+\frac{1}{4}\right)^{-s} =\frac{V_{\mathbb{H}^5}}{3\pi^{\frac{5}{2}}}\, 2^{2s-11}(8s-17)\,\frac{\Gamma\big(s-\frac{5}{2}\big)}{\Gamma(s)}\, , \qquad s>\frac{5}{2}\ .
\end{equation}
 By analytic continuation  to $s\rightarrow 0$, the free energy \eqref{freeci} arises as the coefficient of $-s$. This reproduces  \eqref{Fcerocinco}.
  
 In conclusion, we  find a  $\log\rho_0$ term in the free energy for the CFT on $\mathbb{H}^5$
 with a precise coefficient. The result is different from $\mathbb{S}^5$, suggesting the presence of a conformal anomaly, whose origin must be boundary contributions to the trace
 of the stress tensor.

\paragraph{$\mathbb{H}^7$}: 

The conformal mass is now $M^2=-\frac{35}{4}$. From  \eqref{calor}, \eqref{heathyp}, we now obtain

\begin{equation}
F_{(0,7)}=-\frac{1}{2}\frac{V_{\mathbb{H}^7}}{(4\pi)^{\frac{7}{2}}}\int_{\delta}^\infty \frac{dt}{t^{\frac{9}{2}}}(1+2t+\frac{16}{15}t^2)\, e^{-\frac{t}{4}}\ .
\end{equation}
Subtracting the power-law divergent terms in $\delta $  as above, we now find

\begin{equation}
\label{Fcerosiete}
F_{(0,7)}=\frac{367}{1935360}\frac{1}{\epsilon} = \frac{367}{1935360}\log \rho_0\ .
\end{equation}

We can again reproduce the same result from the explicit sum over eigenvalues. Now the starting point is

\begin{equation}
F_{(0,7)}=\frac{V_{\mathbb{H}^7}}{480\pi^4}\int_0^{\infty}d\lambda\, (4+5\lambda+\lambda^2)\,\log(\lambda+\frac{1}{4})\, .
\end{equation}
Using $\zeta $ function regularization, we recover the result \eqref{Fcerosiete} above. Thus we expect that the CFT on $\mathbb{H}^7$ should also have  a  conformal anomaly
originating from boundary terms.

%%%%%%%%%%%%%%%%%%%%%%
\subsubsection{$\mathbb{S}^2\times \mathbb{H}^3$}

From the general formula \eqref{freegen}, for this space we have

\begin{equation}
F_{(2,3)}=\frac{1}{2}\sum_{l=0}^{\infty}(2l+1)\,\int_{1}^{\infty}d\lambda\,  \Phi_{(3)}\,\log\Big(\lambda-1+\frac{(2l+1)^2}{4}\Big)\ .
\end{equation}
Using the explicit form of the density of eigenvalues given in \eqref{eigendens} and shifting the integration variable $\lambda\rightarrow \lambda+1$, we obtain

\begin{equation}
F_{(2,3)}=\frac{V_{\mathbb{H}^2}}{8\pi^2}\sum_{l=0}^{\infty}(2l+1)\,\int_{0}^{\infty} d\lambda  \sqrt{\lambda} \,\log\Big(\lambda+\frac{(2l+1)^2}{4}\Big)\ .
\end{equation}
We can now introduce Schwinger's proper-time parameter and write 

\begin{equation}
F_{(2,3)}=-\frac{V_{\mathbb{H}^2}}{8\pi^2}\sum_{l=0}^{\infty}(2l+1)\,\int_{\delta}^{\infty} \frac{dt}{t} \int_{0}^{\infty}  d\lambda  \sqrt{\lambda} \,e^{-t\Big(\lambda+\frac{(2l+1)^2}{4}\Big)}\ .
\end{equation}

Computing the integral over $\lambda$, we find

\begin{equation}
\label{F23bare}
F_{(2,3)}=-\frac{V_{\mathbb{H}^3}}{2\,(4\pi)^{\frac{3}{2}} }\int_{\delta}^{\infty} \frac{dt}{t^{\frac{5}{2}}} \sum_{l=0}^\infty (2l+1)\,e^{-\frac{(2l+1)^2}{4}\, t}\ .
\end{equation}
This  recovers the heat kernel formula \eqref{calor}. The integral contains non-physical divergences in the $\delta\rightarrow 0$ limit. The finite part of the integral is obtained by
an appropriate subtraction, by defining the regularized free energy as follows:

\begin{equation}
F_{(2,3)}=-\frac{V_{\mathbb{H}^3}}{2\,(4\pi)^{\frac{3}{2}} }\int_{0}^{\infty} \frac{dt}{t^{\frac{5}{2}}} \sum_{l=0}^\infty (2l+1)\, \left(e^{-\frac{(2l+1)^2}{4}\, t} - 1 + \frac{(2l+1)^2}{4}\, t\right)\ .
\end{equation}
Computing the  integral over $t$ (or, alternatively, keeping the finite part in the $\delta\rightarrow 0$ limit of \eqref{F23bare}) gives 

\begin{equation}
F_{(2,3)}=-\frac{V_{\mathbb{H}^3}}{96\pi}\,\sum_{l=0}^\infty (2l+1)^4\, .
\end{equation}
The sum can be computed using zeta function regularization, using

\begin{equation}
 \sum_{l=0}(2l+1)^{-s}= \left(1-2^{-s}\right) \zeta (s)\ .
\end{equation}
Since $\zeta(-4)=0$, we obtain

\begin{equation}
\label{Fdostres}
F_{(2,3)}=0\ .
\end{equation}
It should be noted that this result strictly holds for the coefficient of the $1/\epsilon $ pole in $V_{\mathbb{H}^3}=-2\pi/\epsilon $.
A finite part of $O(\epsilon^0) $ is in general expected, though
it is much more subtle to calculate and it is likely to be non-universal.
A possible approach to compute the $O(\epsilon^0) $ term in  full-fledged dimensional regularization is by starting with
\eqref{freegen} with $\Phi_{(3-\epsilon)}$ given by \eqref{degi}.
This  leaves a finite  number containing combinations  of $\zeta'(-4) $, $\gamma_E$  and derivatives of $\Gamma$ functions, which indeed suggests strong dependence
on the regularization scheme.

%%%%%%%%%%%%%%%%%%%%%%%%%%%%%%%%%
\subsubsection{$\mathbb{S}^2\times \mathbb{H}^5$}

Using \eqref{freegen}, here we get

\begin{equation}
F_{(2,5)}=\frac{1}{2}\sum_{l=0}^{\infty}(2l+1) \int_{4}^{\infty} d\lambda\  \Phi_{(5)}\,\log\Big(\lambda-4+\frac{(2l+1)^2}{4}\Big)\ ,
\end{equation}
with $\Phi_{(5)}$ given in \eqref{eigendens}.
We can follow the same steps as before. After shifting $\lambda\rightarrow\lambda+4$, and upon introducing a proper-time parameter, the $\lambda$ integral is easily done, leading to

\begin{equation}
F_{(2,5)}=-\frac{V_{\mathbb{H}^5}}{6\,(4\pi)^{\frac{5}{2}} }\int_{\delta}^{\infty} \frac{dt}{t^{\frac{7}{2}}} \sum_{l=0}(2l+1)\,(3+2t)\,e^{-\frac{(2l+1)^2}{4}\, t}\ .
\end{equation}
Computing the $t$ integral keeping the physical finite terms in the $\delta\rightarrow 0$ limit
and performing the $l$ sum as above, we find

\begin{equation}
F_{(2,5)}=\frac{V_{\mathbb{H}^5}}{11520\pi^2} \sum_{l=0}^\infty \left(3(2l+1)^6 -20(2l+1)^4\right)\ .
\end{equation}
In $\zeta $-function regularization, this is proportional to a linear combination of $\zeta(-4)$ and $\zeta(-6)$. Therefore

\begin{equation}
F_{(2,5)} =0 \ .
\end{equation}
As in the previous case, this result strictly holds for the coefficient of the $1/\epsilon $ pole in $V_{\mathbb{H}^5}=\pi^2/\epsilon $.
In dimensional regularization there is a residual $O(\epsilon^0)$ finite part  which appears to be strongly sensitive to the regularization scheme.  

%%%%%%%%%%%%%%%%%%%%%%%%%%%%%%%%%%%%%
\subsubsection{$\mathbb{S}^4\times \mathbb{H}^3$}

For this space,  \eqref{freegen} becomes

\begin{equation}
F_{(4,3)}=\frac{1}{12}\sum_{l=0}^{\infty}(l+1)(l+2)(2l+3) \int_{1}^{\infty}d\lambda \  \Phi_{(3)}\,\log\Big(\lambda-1+\frac{(2l+3)^2}{4}\Big)\ .
\end{equation}
Using the expression for the eigenvalue density $ \Phi_{(3)}$ given in \eqref{eigendens} and shifting $\lambda-1\to\lambda$ we find

\begin{equation}
F_{(4,3)}=\frac{V_{\mathbb{H}^3}}{48\pi^2}\sum_{l=0}^{\infty}(l+1)(l+2)(2l+3) \int_{0}^{\infty}d\lambda\ \sqrt{\lambda}\,\log\Big(\lambda+\frac{(2l+3)^2}{4}\Big)\ .
\end{equation}
Just as in the previous cases, upon introducing a proper-time parameter, the  integral over $\lambda$ is easily done, giving

\begin{equation}
F_{(4,3)}=-\frac{V_{\mathbb{H}^3}}{24\,(4\pi)^{\frac{3}{2}} }\int_{\delta}^{\infty} \frac{dt}{t^{\frac{5}{2}}} \sum_{l=0}(l+1)(l+2)(2l+3)\,e^{-\frac{(2l+3)^2}{4}\, t}\ .
\end{equation}
Keeping the finite term in the $\delta\rightarrow 0$ expansion of the integral, we find

\begin{equation}
F_{(4,3)}=\frac{V_{\mathbb{H}^3}}{4608\pi}\ \sum_{l=0}^\infty \left((2l+1)^4 -(2l+1)^6\right)\ .
\end{equation}
Using $\zeta $ function regularization, like in the previous case, we find

\begin{equation}
F_{(4,3)}=0\ ,
\end{equation}
for the coefficient of the $1/\epsilon$ pole (as in the two previous cases,  we omit the calculation of the much more subtle $O(\epsilon^0)$ term).

%%%%%%%%%%%%%%%%%%%%%%%%%%%%%%%%%%%%%%%%
\subsection{Spaces of the form $\mathbb{S}^{2n+1}\times \mathbb{H}^{2k+1}$}
%%%%%%%%%%%%%%%%%%%%%%%%%%%%%%%%%%%%%%%%

We have already studied one subclass of these spaces, namely the case $n=0$
corresponding to $\mathbb{S}^{1}\times \mathbb{H}^{2k+1}$.
We found that the free energy exhibits a boundary anomaly and it is proportional to $1/\epsilon =\log\rho_0 $. The free energy does not match the free energy of  $\mathbb{S}^{2k+2}$, but for $\mathbb{S}^{1}\times \mathbb{H}^{2k+1}$  there is a thermodynamic interpretation
by which one can identify  the  entanglement entropy and
and check that it matches with the entanglement entropy on $\mathbb{S}^{2k+2}$.
For the cases with $n>0$, there is no thermodynamic interpretation as one does
not have a thermal circle. However, like in the $n=0$ case, the free energy on
the spaces $\mathbb{S}^{2n+1}\times \mathbb{H}^{2k+1}$ also exhibits a logarithmic
IR divergent term $\log\rho_0$, which, through the UV/IR connection, indicates the presence of a boundary conformal
anomaly.
In what follows, we will compute it for the case $\mathbb{S}^{3}\times \mathbb{H}^{3}$, which is the example of lowest dimension in this class and already
illustrates the main features.

\subsubsection{  $\mathbb{S}^{3}\times \mathbb{H}^{3}$}

Using the heat kernel (for coincident points) for $\mathbb{S}^3$ and adding the mass coming from the conformal coupling to curvature, we find that the
free energy is

\begin{equation}
F_{(3,3)}=- \frac{V_{\mathbb{H}^3}V_{\mathbb{S}^3}}{32\pi^{\frac{7}{2}} } \sum_{n=1}^\infty \int_{\delta}^\infty \frac{dt}{t^{\frac{5}{2}}} n^2 e^{-n^2 t}\ .
\end{equation}
Computing the finite part of the integral and the sum  (using $\zeta$ function regularization), we obtain

\begin{equation}
F_{(3,3)}=\frac{V_{\mathbb{H}^3}V_{\mathbb{S}^3}}{6048\pi^3}\ .
\end{equation}
Using that $V_{\mathbb{S}^3}=2\pi^2$ and that $V_{\mathbb{H}^3}=-\frac{2\pi}{\epsilon}$, we find

\begin{equation}
\label{F33}
F_{(3,3)}=-\frac{1}{1512}\, \frac{1}{\epsilon}\ .
\end{equation}
As in the case $ \mathbb{H}^{3}$ discussed in section 3.3.1, the presence of an $\epsilon $ pole ($1/\epsilon =\log\rho_0$) suggests that the CFT on  $\mathbb{S}^{3}\times \mathbb{H}^{3}$ has a boundary conformal anomaly 
Note that the coefficients of the IR and UV logarithmic terms in $F_{(3,3)}$ and $F_{(6,0)}$ differ by a factor $1/2$.
The CFT in both spaces 
$\mathbb{S}^{3}\times \mathbb{H}^{3}$ and $\mathbb{S}^{6}$ has a conformal anomaly but the origin is different.
The space $\mathbb{S}^{3}\times \mathbb{H}^{3}$ has vanishing Weyl tensor and
vanishing Euler characteristic, since $\chi (\mathbb{S}^{3})=0$. Thus the trace of the stress tensor can only receive contributions from boundary terms.  We have already seen  this feature explicitly in the $\mathbb{H}^{3}$ example in \eqref{traso}.
We will return to this in section \ref{strongcoupling}.

As a check, let us now compute $F_{(3,3)}$ by the alternative method of explicitly
summing over eigenvalues. From \eqref{freegen}, we have

\begin{equation}
F_{(3,3)}=\frac{1}{2}\sum_{l=0}^{\infty}(l+1)^2\,\int_{1}^{\infty} d\lambda \,  \Phi_{(3)}\,\log\Big(\lambda-1+(l+1)^2\Big)\ .
\end{equation}
Substituting $\Phi_{(3)}$ given in \eqref{eigendens} and upon a shift $\lambda\rightarrow \lambda-1$ in the integration variable, we find

\begin{equation}
F_{(3,3)}=\frac{V_{\mathbb{H}^3}}{8\pi^2}\sum_{l=1}^{\infty}l^2\,\int_{0}^{\infty}d\lambda\,   \sqrt{\lambda}\,\log\Big(\lambda+l^2\Big)\ .
\end{equation}
The integral can be regulated by considering 
\begin{equation}
\int_{0}^{\infty}d\lambda  \sqrt{\lambda}\,\Big(\lambda+l^2\Big)^{-s} =\frac{l^{3-2s} \sqrt{\pi} \Gamma(s-3/2)}{2\Gamma(s)}\ ,
\end{equation}
and extracting the linear term in $s$ in the expansion in powers of $s$.
 Then, computing the sum over $l$ using $\zeta$ function regularization, 
 we reproduce the result \eqref{F33}.

%%%%%%%%%%%%%%%%%%%%%%%%%%%%%%%

\section{Strongly coupled fields on $\mathbb{S}^a\times \mathbb{H}^b$}\label{strongcoupling}

Let us consider a general CFT in $d=a+b$ dimensions at strong coupling, admitting a gravity dual as a solution to the Einstein-Hilbert action in $D=a+b+1$ dimensions with non-zero cosmological constant,

\begin{equation}
\label{EH}
S_{EH}=-\frac{1}{16\pi G_N}\int d^Dx\,\sqrt{g}\,\Big(R-2\Lambda\Big)\,,\qquad \Lambda=-\frac{(D-1)(D-2)}{2}\, .
\end{equation}
A vacuum solution is $AdS_D$,  which can be written in the following coordinates,

\begin{equation}
\label{AdSp}
ds^2=\frac{dr^2}{(1+r^2)}+r^2\,ds_{\mathbb{S}^a}^2+(1+r^2)\,ds_{AdS_b}^2\,.
\end{equation}
As $r\rightarrow \infty$, the above metric is asymptotic to 

\begin{equation}
ds^2=\frac{dr^2}{r^2}+r^2\,(ds_{\mathbb{S}^a}^2+ds_{AdS_b}^2)\, .
\end{equation}
Therefore, \eqref{AdSp} describes  $AdS_D$ space with boundary $\mathbb{S}^a\times AdS_b$. Thus, this space is the natural candidate to support the holographic dual to the CFT on $\mathbb{S}^a\times AdS_b$. We can now Wick rotate the $AdS_b$ into $\mathbb{H}^b$ to find the holographic dual of a generic CFT$_{a+b}$ on $\mathbb{S}^a\times \mathbb{H}^b$.

\subsection{Free energy from gravity}

Let us evaluate the on-shell action on our background. Using the value of the scalar curvature $R=-D(D-1)$, we find

\begin{equation}
S_{EH}^{\rm os}=\frac{(a+b)}{8\pi\, G_N}\int d^{a+b+1}x\,\sqrt{g}=\frac{(a+b)}{8\pi\, G_N}\,\Big(\int_{\mathbb{S}^a}\sqrt{g_{\mathbb{S}^a}}\Big)\, \Big(\int_{\mathbb{H}^b}\sqrt{g_{\mathbb{H}^b}}\Big)\,  \int_0^{r_0} dr \,r^a\,(1+r^2)^{\frac{b-1}{2}} ,
\end{equation}
where we introduced a  cut-off $r_0$ in the radial coordinate. Then

\begin{equation}
S_{EH}^{\rm os}=\frac{V_{\mathbb{S}^a}\, V_{\mathbb{H}^b}}{8\pi\, G_N} \,\frac{(a+b) }{(a+1)}\, r_0^{a+1}\, \,_2F_1(\frac{a+1}{2},\frac{1-b}{2},1+\frac{a+1}{2};-r_0^2)\, .
\end{equation}

As it is well-known, in order to have a well-defined variational problem, the action should be supplied by the Gibbons-Hawking surface term, which must be then evaluated on-shell to compute the holographic free energy:

\begin{equation}
S^{\rm os}_{GH}=-\frac{1}{8\pi G_N}\int_{\partial} \sqrt{\gamma}\,\Theta\, .
\end{equation}
Here $h_{ab}$ is the induced metric on the boundary and $\Theta$ is the extrinsic curvature. Note that a vector normal to the boundary is 

\begin{equation}
n=\sqrt{1+r^2}\partial_r\, ,
\end{equation}
while the boundary metric is

\begin{equation}
ds_{\gamma}^2=r^2ds_{\mathbb{S}^a}^2+(1+r^2)ds_{\mathbb{H}^b}^2\, .
\end{equation}
Thus 

\begin{equation}
\sqrt{\gamma} = r^a\,(1+r^2)^{\frac{b}{2}}\,\sqrt{g_{\mathbb{S}^a}}\,\sqrt{g_{\mathbb{H}^b}}\, .
\end{equation}
Therefore, the Gibbons-Hawking term evaluated on the cut-off surface is\footnote{As a check, note that $n=\sqrt{1+r^2}\,\partial_r$. Therefore $n^r=\sqrt{1+r^2}$. Then $\Theta_{\mu\nu}=-\gamma_{\mu}^{\rho}\nabla_{\rho} n_{\nu}$, so $\Theta_{ab}=-\gamma_{a}^{c}\nabla_{c} n_{b}$, where latin indices stand for boundary indices. In turn, since $\nabla_{\rho}n_{\nu}=\partial_{\rho}n_{\nu}+\Gamma^{\alpha}_{\nu\rho}n_{\alpha}$, we have $\nabla_{c}n_{b}=\partial_{c}n_{b}+\Gamma^{r}_{bc}n_{r}=\Gamma^r_{bc}n_r=\frac{1}{2}g^{rr}\partial_r g_{bc}\,n_r$. Therefore ($g_{ab}=\gamma_{ab}$) $\Theta^a\,_{b}=\frac{\sqrt{1+r^2}}{2}\,\gamma^{ac}\partial_r \gamma_{cb}$
Then, as a matrix

\begin{equation}
\Theta^a\,_b=\left(\begin{array}{c | c} \frac{\sqrt{1+r^2}}{r}\,\unity_{a\times a} & \\ \hline & \frac{r}{(1+r^2)^{\frac{1}{2}}}\,\unity_{b\times b}\end{array}\right)\, .
\end{equation}
From here it follows that  $\Theta=\Theta^b\,_b=\frac{a+(a+b)r^2}{r\,\sqrt{1+r^2}}$.}

\begin{equation}
S_{GH}^{\rm os} = -\frac{V_{\mathbb{S}^a}\, V_{\mathbb{H}^b}}{8\pi G_N}  \,r_0^{a-1}\,(1+r_0^2)^{\frac{b-1}{2}}\,(a+(a+b)r_0^2)\,.
\end{equation}

In addition, in order to implement holographic renormalization, we need to add  counterterms (to be evaluated on-shell as well). Up to dimension $D=6$, they are given by \cite{Emparan:1999pm}

\begin{equation}
S_{CT}^{\rm os}=\frac{1}{8\pi\,G_N}\int_{\partial} \sqrt{\gamma}\,\Big[(D-2)+\frac{1}{2(D-3)}\,\mathcal{R}+\frac{1}{2(D-5)(D-3)^2}\,\Big(\mathcal{R}_{ab}\mathcal{R}^{ab}-\frac{D-1}{4(D-2)}\,\mathcal{R}^2\Big)\Big]\, ,
\end{equation}
where $\mathcal{R}_{ab}$ is the Ricci curvature of the boundary metric and $\mathcal{R}$ is its scalar curvature
(the second and third counterterms are strictly needed only for $D>3$ and
$D>5$, respectively). For the space $\mathbb{S}^a\times \mathbb{H}^b$, one has

\begin{equation}
\mathcal{R}=\frac{a(a-1)}{r_0^2}-\frac{b(b-1)}{1+r_0^2}\ ,\qquad \mathcal{R}_{ab}\mathcal{R}^{ab}=\frac{a(a-1)^2}{r_0^4}+\frac{b(b-1)^2}{(1+r_0^2)^2}\, .
\end{equation}
Using these ingredients, we are in place to holographically compute the free energy corresponding to strongly coupled CFT's  dual on $\mathbb{S}^a\times \mathbb{H}^b$ as

\begin{equation}
F_{(a,b)}^{\rm holo}=S_{EH}^{\rm os}+S_{GH}^{\rm os}+S_{CT}^{\rm os}\, .
\label{renoac}
\end{equation}
The case $a=1$ has been already considered in \cite{Emparan:1999pm}. Thus, in the following we will discuss the remaining cases.

\subsubsection{$\mathbb{S}^{2n+1}\times \mathbb{H}^{2k}$}

These cases contain no logarithmic divergent $\log r_0$ term. Expanding the renormalized action \eqref{renoac} at large $r_0$, we find that the leading term is cut-off independent and reads (recall that here $a=2n+1$, $b=2k$)

\begin{equation}
F_{(a,b)}^{\rm holo}=\frac{V_{\mathbb{S}^a}\, V_{\mathbb{H}^b}}{8\pi G_N}\,\frac{(a+b)\,\Gamma\Big(\frac{3+a}{2}\Big)\,\Gamma\Big(\frac{-a-b}{2}\Big)}{(1+a)\,\Gamma\Big(\frac{1-b}{2}\Big)}\, .
\end{equation}
Since $b=2k$, there is no logarithmic IR divergent term in the volume of  $\mathbb{H}^b$ (see appendix \ref{appVolHb}). 
Substituting the value of the volumes of  $\mathbb{S}^a$ and $\mathbb{H}^b$ , this becomes ($D=a+b+1$)

\begin{equation}
F_{(2n+1,2k)}^{\rm holo}=-\frac{\pi^{\frac{D-3}{2}}\,\Gamma\big(\frac{3-D}{2}\big)}{4G_N}=F_{(2n+1+2k,0)}^{\rm holo}\, ,
\end{equation}
where in the last step we have used \eqref{FSphere}. This shows that, just as in the free scalar model of section 3, the free energy for a strongly coupled CFT on $\mathbb{S}^{2n+1}\times \mathbb{H}^{2k}$ equals that on $\mathbb{S}^{2n+1+2k}$, thus showing the absence of conformal  anomalies --in particular, the absence of boundary conformal anomalies.

\subsubsection{$\mathbb{S}^{2n}\times \mathbb{H}^{2k}$}

Let us now consider   cases of even total dimension, concentrating on spaces of the form $\mathbb{S}^{2n}\times \mathbb{H}^{2k}$. Being an even-dimensional space, one should in general expect the presence of bulk conformal anomalies, showing up in the free energy through a logarithmic term in $\log r_0$. Starting with the case of total dimension 4, a straightforward application of \eqref{renoac} yields (we only quote the logarithmic term)

\begin{equation}
F_{(4,0)}^{\rm holo} = F_{(2,2)}^{\rm holo}= \frac{\pi}{2G_N}\,\log r_0\, , \qquad F_{(0,4)}^{\rm holo}= \frac{\pi}{4G_N}\,\log r_0\, .
\label{cuacero}
\end{equation}
The $\mathbb{S}^4$ and $\mathbb{H}^4$ cases were originally computed in \cite{Emparan:1999pm}, being both formally the same up to a factor of two due to the volume ratio $V_{\mathbb{S}^4}/V_{\mathbb{H}^4}=2$. 

Let us now move to the case of total dimension 6. We now find

\begin{equation}
F_{(6,0)}^{\rm holo} =F_{(4,2)}^{\rm holo}=F_{(2,4)}^{\rm holo}=-\frac{\pi^2}{4G_N}\,\log r_0\, ,\qquad F_{(0,6)}^{\rm holo}=-\frac{\pi^2}{8G_N}\,\log r_0\,.
\label{seiscero}
\end{equation}
The cases of the $\mathbb{S}^6$ and the $\mathbb{H}^6$ were also computed in  \cite{Emparan:1999pm}, finding that they are formally identical up to a relative factor of 2 originating from the volume factors as described above. On the other hand, the remaining cases yield a free energy equal to that of the $\mathbb{S}^6$.

In order to understand these results, note first that the boundary of $\mathbb{H}^{2k}$ is $\mathbb{S}^{2k-1}$, an odd-dimensional space.
While in \cite{Dowker:1989ue,Herzog:2015ioa,Solodukhin:2015eca} conformal boundary anomalies have also been proposed for odd-dimensional boundaries, here we are finding that, at least for
our spaces, there are no conformal boundary anomalies when the dimension of the boundary is odd (see section \ref{discussion} for further comments on this). In turn, the bulk conformal anomaly must come from the $A$-type
anomaly, and hence the ratio of free energies must be proportional to the ratio of Euler numbers.
The Euler number of the non-compact hyperbolic space 
can be defined as usual by including a suitable boundary term in the definition. 
This gives $\chi(\mathbb{H}^{2k})=1$.\footnote{
This is also consistent with  our regularization for $V_{\mathbb{H}^{2k}}$ 
and a simple application of the Gauss-Bonnet formula  \cite{Hopf,Weil}, $V_{\mathbb{S}^{2k}}\chi (\mathbb{H}^{2k}) = (-1)^k 2V_{\mathbb{H}^{2k}}$.}
 Therefore  $\chi(\mathbb{S}^{2n}\times \mathbb{H}^{2k})=\chi(\mathbb{S}^{2n+2k})$ as long as $n\ne 0$, while $\mathbb{\chi}(\mathbb{H}^{2k})=\frac{1}{2}\chi(\mathbb{S}^{2k})$, thus precisely matching the pattern of free energies \eqref{cuacero}, \eqref{seiscero} which we have found.

\subsubsection{$\mathbb{S}^{2n}\times \mathbb{H}^{2k+1}$}

Let us briefly comment on the case $\mathbb{S}^{2n}\times \mathbb{H}^{2k+1}$. The particular case $n=0$, $k=1$ was computed in  \cite{Emparan:1999pm}, where it was shown to vanish. This result holds for all spaces of the form $\mathbb{H}^{2k+1}$,
moreover, for all spaces of the form
$\mathbb{S}^{2n}\times \mathbb{H}^{2k+1}$. This can be seen as follows.
After expanding \eqref{renoac} in powers of $r_0$, we find that there is no finite (nor any logarithmic) term. For instance, for the case $\mathbb{S}^2\times \mathbb{H}^3$, one finds

\begin{equation}
F_{(2,3)}^{\rm holo}=\frac{V_{\mathbb{S}^2}V_{\mathbb{H}^3}}{32\pi G_N}\frac{1}{r_0}-\frac{V_{\mathbb{S}^2}V_{\mathbb{H}^3}}{768\pi G_N}\frac{1}{r_0^3}+\cdots\rightarrow 0\, .
\end{equation}
Note that inside $V_{\mathbb{H}^{2k+1}}$ there is a hidden $\log\frac{R}{\rho_0}$, where $\rho_0$ is a cutoff in the hyperbolic space $\mathbb{H}^{2k+1}$ (this corresponds to the pole in DREG, see appendix A). But all terms vanish as $r_0\to\infty$. Remarkably, this is entirely consistent with our findings for conformal free scalars on $\mathbb{S}^{2n}\times \mathbb{H}^{2k+1}$, where the free energy  in these spaces was found to vanish, at least for the coefficient of $V_{\mathbb{H}^{2k+1}}$ in dimensional regularization.

It would be interesting to understand if there is a finite remnant, perhaps originating from  surface terms on the boundary of $\mathbb{H}^{2k+1}$,  and whether this matches with the free energy \eqref{FSphere} of $\mathbb{S}^{2n+2k+1}$.

\subsubsection{$\mathbb{S}^{2n+1}\times \mathbb{H}^{2k+1}$}

Here we consider as an example $\mathbb{S}^{3}\times \mathbb{H}^{3}$. Using our expressions above, we find that

\begin{equation}
F_{(3,3)}^{\rm holo} = \frac{V_{\mathbb{H}^3}\,V_{\mathbb{S}^3}}{32\pi G_N}\,.
\end{equation}
Recall, nevertheless, that $V_{\mathbb{H}^3}$ contains a $\log\frac{R}{\rho_0}$ term (or an $\epsilon $ pole, in DREG). Therefore, making use of our formula \eqref{VolHb} and the volume of the 3-sphere \eqref{VolSa}, we find

\begin{equation}
F_{(3,3)}^{\rm holo} = -\frac{\pi^2}{8G_N}\,\log \rho_0\ .
\label{trestres}
\end{equation}
Note that this is neither the free energy on $\mathbb{S}^6$,  $\mathbb{S}^4\times \mathbb{H}^2$, $\mathbb{S}^2\times \mathbb{H}^4$, nor the free energy on  $\mathbb{H}^6$, thus explicitly showing the presence of a conformal anomaly.
We recall that there is an $A$-anomaly on the spaces $\mathbb{S}^{2n}\times \mathbb{H}^{6-2n}$. On the other hand, the $A$-anomaly vanishes on $\mathbb{S}^{2n+1}\times \mathbb{H}^{2k+1}$, since the Euler characteristic is zero,
but there can still be a contribution to the conformal anomaly  from
boundary terms (see \cite{Solodukhin:2015eca} for a general construction).

The logarithmic dependence on the scale suggests that the conformal anomaly produced by boundary terms can be read from the coefficient of the log term in
\eqref{trestres}. It is interesting that this coefficient differs from the $A$ anomaly
coefficient on $\mathbb{S}^{6}$, given
in
\eqref{seiscero}, by a factor $1/2$, and that the same relative factor appears
in the ratio of the coefficients of IR and UV log terms in  $F_{(3,3)}$ and $F_{(6,0)}$ computed at weak coupling for a free conformal scalar (see \eqref{F33}, \eqref{segunda}). 
This could be a consequence of the form of the boundary contribution to the anomaly. From the expressions for two-dimensional and four-dimensional boundaries discussed in \cite{Solodukhin:2015eca}, one may guess that such boundary anomaly 
involves either the Euler number of the boundary or the Weyl tensor and trace-free part of the extrinsic curvature (c.f. \eqref{traso}). Since the latter two tensors vanish in our case, the boundary anomaly would be given by a general expression of the form $\int \langle T\rangle = c_{\partial} \chi(\partial \mathcal{M})$ for some coefficient $c_{\partial}$. Our findings suggest that, at least as long as $n\ne 0$, $c_{\partial}$ is proportional to the $a$ central charge with some universal coefficient whose origin would be very interesting to clarify.

\section{Discussion}\label{discussion}

Boundary conformal anomalies have been comparatively poorly studied with respect to their bulk counterparts. Indeed, to the best of our knowledge, there is no comprehensive study of these in arbitrary dimension. In this paper we have introduced an interesting class of spaces, namely $\mathbb{S}^a\times \mathbb{H}^b$, conformally related to $\mathbb{S}^{a+b}$ where boundary anomalies  play an important role. 

We have studied a free conformal scalar as well as strongly coupled CFT's (the latter through holography) on $\mathbb{S}^a\times \mathbb{H}^b$. 
The case $a=1$ is somewhat special, as it permits an interpretation in terms of entanglement entropy across a $b-2$-dimensional sphere. Through this connection it is possible to argue that the relation between $F_{(1+b,0)}$ and $F_{(1,b)}$ is precisely given by \eqref{relF}. 
It is worth noting that  $F_{(1+b,0)}-F_{(1,b)}=-\beta\partial_{\beta}F_{(1,b)}\equiv \Delta_b$, when evaluated at $\beta=2\pi$, measures the   total conformal anomaly $\Delta_b $, coming in principle both from bulk and boundary contributions (see \textit{e.g.} \cite{Beccaria:2017dmw}).
Recall now that $\Delta_b$ is zero for even $b=2k$, while, as argued in \cite{Herzog:2015ioa}, $\Delta_b$ contains boundary contributions for odd $b=2k+1$. The boundary anomaly is supported on the even-dimensional $\mathbb{S}^{2k}$ at the boundary of $ \mathbb{H}^{2k+1}$. 

More generally, consider the families of spaces  $\mathbb{S}^{2n+1}\times \mathbb{H}^{2k}$ and $\mathbb{S}^{2n+1}\times \mathbb{H}^{2k+1}$.
The bulk conformal anomaly vanishes on these spaces because they have zero Euler characteristic and vanishing Weyl tensor (so both $a$ and $c$ bulk anomaly contributions vanish).
For the first family of spaces, $\mathbb{S}^{2n+1}\times \mathbb{H}^{2k}$, our results, both at weak coupling and strong coupling,  show that $F_{(2n+1,2k)}=F_{(2n+1+2k,0)}$, implying that also the boundary conformal anomaly 
vanishes.\footnote{This is consistent with the general expression for the trace of the stress tensor for spaces with odd-dimensional boundaries given in \cite{Solodukhin:2015eca}, which is expressed in terms of the Weyl tensor and the trace-free part of the extrinsic curvature, both  vanishing  when the boundary is $\mathbb{S}^{2k-1}$.}
However, for the second family of spaces, we find $F_{(2n+1,2k+1)}\ne F_{(2n+2k+2,0)}$.
This shows that, at least in the class of spaces  $\mathbb{S}^{2n+1}\times \mathbb{H}^{b}$, boundary conformal anomalies only appear when the boundary  space
$\mathbb{S}^{b-1}$ has even dimension.
 
In the case of $\mathbb{H}^3$, we related our result to the general expression 
for the boundary contribution to the trace of the stress tensor, given in \cite{Solodukhin:2015eca}.
We stress that  it is the coefficient of an IR logarithmic divergence what is related to the conformal anomaly. This is related to the underlying UV/IR connection discussed in appendix A (see also \cite{Casini:2011kv,Herzog:2015ioa}). 
%aqui
This implies that a short-distance cutoff $\delta $ on a conformally related geometry is equivalent to an IR cutoff $\rho_0$ on $\mathbb{H}^3$, $\delta\sim 1/\rho_0$.
In additon, assuming this connection, we used the formula in \cite{Solodukhin:2015eca}
to compute the trace of the stress tensor, finding a perfect matching
with the prediction coming from the coefficient of the IR logarithmic divergence.
It would be extremely interesting to 
undertake a general analysis of the boundary anomalies extending the work of \cite{Solodukhin:2015eca}, perhaps leading to a prediction of the boundary anomalies found here, presumably in terms of the central charges of the the CFT.

 As another example of this, we have  studied the case of $\mathbb{S}^3\times \mathbb{H}^3$, where we have explicitly seen the appearance of the boundary anomaly. 
 Interestingly, we have seen that the ratio of coefficients of the IR and UV logarithmic terms in $F_{(3,3)}$ and $F_{(6,0)}$ is $1/2$, both for the weak coupling computation as well as for the holographic computation. 
 We have also computed  $F_{(5,5)}$, finding an IR logarithmic
 term whose coefficient is 1/2 the coefficient in the UV logarithmic term of $F_{(10,0)}$. It would be very interesting to prove such patterns, compare them to holography and understand them from the general form of the trace of the stress tensor.

The odd-dimensional cases $\mathbb{S}^{2n}\times \mathbb{H}^{2k+1}$ remain puzzling. Since they are free from bulk conformal anomalies, they may only suffer from boundary anomalies. Indeed, the case $n=0$, $k=1$ allowed us to explicitly test this by matching our result to the prediction of \cite{Solodukhin:2015eca}. However, for $n\ne 0$, we seem to find that there is no logarithmic term (i.e. no pole in DREG).\footnote{Our results show that there is a scheme where the logarithmic term is absent. This property should be universal in the sense that it still holds under scaling of the cutoff. However, a logarithmic term might in principle appear in another regularization scheme. We thank A. Tseytlin for making this point.} 
The absence of a logarithmic term suggests that there is no boundary anomaly either. In turn, at strong coupling the holographic computation gives a vanishing free energy irrespective of the value of $n$. It would be important to understand these cases at least qualitatively, since, in principle, 
 extra counterterms due to boundary effects might give new non-trival contributions \cite{Graham:1999pm,Takayanagi:2011zk,Miao:2017gyt,Chu:2017aab,Astaneh:2017ghi}.

We also showed that supersymmetric field theories can be defined on  spaces  $\mathbb{S}^{a}\times \mathbb{H}^b$ (see appendix C). It would be extremely interesting
to compute the partition function on $\mathbb{S}^{a}\times \mathbb{H}^b$ by supersymmetric localization for  supersymmetric gauge theories in various dimensions. In particular, in
the large $N$ limit, the localization results may be directly compared with our results for the holographic
free energy.\footnote{Examples of localization on  hyperbolic spaces can be found in \cite{David:2016onq,Assel:2016pgi,Cabo-Bizet:2017jsl}.}

\section*{Acknowledgements}

We are especially grateful to A.~Tseytlin for many useful comments that led to important corrections on a preliminary version of this work.
We would also like to thank in particular T.~Nishioka for  important explanations
and  remarks, and A.~Bourget, C.~Hoyos, E.~Teste and A.~Yarom  for useful discussions.
D.R-G is partly supported by the Ramon y Cajal grant RYC-2011-07593 as well as the EU CIG grant UE-14-GT5LD2013-618459, the Asturias Government grant FC-15-GRUPIN14-108 and Spanish Government grant MINECO-16-FPA2015-63667-P. J.G.R. acknowledges financial support from projects  FPA2013-46570  
and  MDM-2014-0369 of ICCUB (Unidad de Excelencia `Mar\'ia de Maeztu').

\begin{appendix}

\section{Regularized volume of $\mathbb{H}^b$
% volume of  $\mathbb{S}^a$ 
and UV/IR connection}\label{appVolHb}

Here we  review the computation of the regularized volume of $\mathbb{H}^b$   \cite{Diaz:2007an,Casini:2011kv,Casini:2010kt}. Let us consider $\mathbb{H}^b$ with metric

\begin{equation}
ds^2=dy^2+\sinh^2y\,ds_{\mathbb{S}^{b-1}}^2=\frac{d\rho^2}{1+\rho^2}
+\rho^2\,ds_{\mathbb{S}^{b-1}}^2\, .
\end{equation}
Following \cite{Diaz:2007an,Casini:2011kv,Casini:2010kt}, we regularize the volume by putting a UV boundary at $\rho_0 $ as

\begin{equation}
V_{\mathbb{H}^b} = 
V_{\mathbb{S}^{b-1}}\,\int_0^{\rho_0} d\rho\,\frac{\rho^{b-1}}{\sqrt{1+\rho^2}}\,.
\end{equation}
Expanding at large $\rho_0$, one finds a logarithmic divergence when
$b$ is odd.
Alternatively, in the limit $\rho_0\to \infty$, the integral can be computed in dimensional regularization. This gives
\begin{equation}
\label{VolHb}
V_{\mathbb{H}^b}=\frac{\Gamma\big(\frac{1-b}{2}\big)}{\pi^{\frac{1-b}{2}}}\,.
\end{equation}
For odd $b$ one can expand as $b=2n+1-\epsilon$ for $\epsilon\rightarrow 0^+$. The leading pole in $\epsilon$ gives the coefficient of the logarithmic term $\log\rho_0$ (see discussions in \textit{e.g.}  \cite{Perlmutter:2013gua}).

We will also make use of the explicit formula for the  volume of an $a$-dimensional sphere:
\begin{equation}
\label{VolSa}
V_{\mathbb{S}^a}=\frac{2 \pi^{\frac{a+1}{2}}}{\Gamma\Big(\frac{a+1}{2}\Big)}\, .
\end{equation}

Let us now discuss the connection between the IR  cutoff on  $ \mathbb{S}^a\times \mathbb{H}^{2k+1}$ and the UV cutoff on $ \mathbb{S}^{a+2k+1}$ induced by the conformal map, 
generalizing the arguments of \cite{Casini:2011kv} for  $ \mathbb{S}^1\times \mathbb{H}^{2k+1}$ to our case. From (\ref{metriga}), we see that a  covariant, short-distance  UV cutoff $\delta_{\mathbb{S}^{2k+1}}^2$ on the sphere is related to a covariant UV cutoff $\delta_{\mathbb{H}^{2k+1}}^2$ on the hyperbolic space by
\begin{equation}
\delta_{\mathbb{S}^{2k+1}}^2 \approx  \frac{1}{ \rho_0^2} \delta_{\mathbb{H}^{2k+1}}^2\ ,
\end{equation}
where we have used $\rho=\sinh y$, $\rho_0\gg 1$.
Thus the  UV momentum cutoff on the sphere $\Lambda \equiv 1/\delta_{\mathbb{S}^{2k+1}}$ is linearly related to  the IR cutoff $\rho_0$ on the volume of $\mathbb{H}^{2k+1}$.
The case $a=0$ has to be treated separately, since the conformal map is different. Consider for example  $\mathbb{H}^{2k+1}$ with metric
\begin{equation}
ds^2_{\mathbb{H}^{2k+1}}= \frac{dx_i dx_i}{(1-x_i x_i)^2}= \frac{d\rho^2}{1+\rho^2}
+ \rho^2 ds^2_{\mathbb{S}^{2k}}\ .
\end{equation}
An IR cutoff at $\rho_0\gg 1$
 implies $r^2= x_ix_i=1-2\delta$, $\delta \sim 1/\rho_0$.
A covariant UV cutoff $\delta_{\mathbb{S}^{2k+1}}^2$ on the sphere is related to a covariant UV cutoff $\delta_{\mathbb{H}^{2k+1}}^2$ on the hyperbolic space by
\begin{equation}
\delta_{\mathbb{S}^{2k+1}}^2 =  \frac{(1-r^2)^2}{(1+r^2)^2} \delta_{\mathbb{H}^{2k+1}}^2
\approx \frac{1}{\rho_0^2} \, \delta_{\mathbb{H}^{2k+1}}^2\ ,
\end{equation} 
leading to the same linear relation between  $\Lambda =1/\delta_{\mathbb{S}^{2k+1}}$ and $\rho_0$.

\section{Free energy on spheres}\label{FinSphere}\label{appHoloFSphere}

In this appendix we include the holographic derivation of the free energy for a CFT on $\mathbb{S}^{D-1}$. To that matter, let us consider euclidean $AdS_D$ with metric 

\begin{equation}
ds^2=\frac{dr^2}{1+r^2}+r^2ds_{\mathbb{S}^{D-1}}^2\,.
\end{equation}
Since the curvature of $AdS_D$ is $R=-D(D-1)$, the on-shell Einstein-Hilbert action is

\begin{eqnarray}
S_{EH}^{\rm os} &=& \frac{V_{\mathbb{S}^{D-1}}}{8\pi\,G_N}\, (D-1) \,\int_0^{r_0}dr\, \frac{r^{D-1}}{\sqrt{1+r^2}}
\nonumber\\
&=&\frac{V_{\mathbb{S}^{D-1}}}{8\pi\,G_N}\,\frac{(D-1)}{D}\,r_0^D\,\,_2F_1(\frac{1}{2},\frac{D}{2},1+\frac{D}{2},-r_0^2)\, .
\label{EHSphere}
\end{eqnarray}

In turn, the induced metric on the boundary is $d\xi^ah_{ab}d\xi^b=r^2 ds_{\mathbb{S}^{D-1}}^2$, while the normal vector to the boundary is

\begin{equation}
n=\sqrt{1+r^2}\, \partial_r\, .
\end{equation}
Thus, the Gibbons-Hawking term is

\begin{equation}
S_{GH}^{\rm os}=-\frac{1}{8\pi G_N}\int_{\partial}\sqrt{1+r^2}\, \partial_r(\sqrt{\gamma})=-\frac{V_{\mathbb{S}^{D-1}}}{8\pi G_N}\, (D-1)\,\sqrt{1+r_0^2}\,r_0^{D-2}\, .
\end{equation}
In order to compute the counterterms, note that, when evaluated at the cut-off surface

\begin{equation}
\mathcal{R}=\frac{(D-1)(D-2)}{r_0^2}\ ,\qquad \mathcal{R}_{ab}=\frac{(D-2)}{r_0^2}\gamma_{ab}\, .
\end{equation}
Hence $\mathcal{R}_{ab}\mathcal{R}^{ab} = \frac{(D-2)^2\,(D-1)}{r_0^4}$. Thus the counterterm piece is

\begin{equation}
S_{CT}^{\rm os}=\frac{V_{\mathbb{S}^{D-1}}}{8\pi\,G_N}\, \Big(-\frac{(D-2)(D-1)}{8\,(D-5)}\,r_0^{D-5}+\frac{(D-2)(D-1)}{2\,(D-3)}\,r_0^{D-3}+(D-2)\,r_0^{D-1}\Big)\, .
\end{equation}
We can now compute the free energy by adding the Einstein-Hilbert, Gibbons-Hawking and counterterm contribution upon expanding for large $r_0$. We find

\begin{equation}
F_{(D-1,0)}^{\rm holo}=\frac{V_{\mathbb{S}^{D-1}}}{8\pi G_N}\,\frac{(D-1)}{D}\frac{\Gamma\Big(\frac{1-D}{2}\Big)\Gamma\Big(\frac{2+D}{2}\Big)}{\sqrt{\pi}}\, .
\end{equation}
 For odd $D$, this formula is to be understood in dimensional regularization,  since there is a $1/\epsilon $ pole associated with a
 logarithmic term (see \cite{Emparan:1999pm}).\footnote{We will also assume minimal holographic renormalization, corresponding to a scheme where $D$-type anomalies are absent. See \cite{Balasubramanian:1999re}, \cite{Huang:2013lhw} for  discussions.} Finally, using the formula \eqref{VolSa} for the volume of the $\mathbb{S}^{D-1}$, we get

\begin{equation}
\label{FSphere}
F_{(D-1,0)}^{\rm holo}=-\frac{1}{8\pi G_N}\,2\pi^{\frac{D-1}{2}}\,\Gamma\Big(\frac{3-D}{2}\Big)\, .
\end{equation}

\section{Supersymmetry on $\mathbb{S}^a\times AdS_b$}\label{appSUSY}

Being conformally related to spheres, one may expect that the family of geometries $\mathbb{S}^a\times AdS_b$ can support supersymmetric theories on them. In this appendix we explicitly study the SUSY of the geometries under consideration (see \cite{Aharony:2015hix} for related developments). Consider the metric

\begin{equation}
ds^2=ds_{\mathbb{S}^a}^2+ds_{AdS_b}^2\, .
\end{equation}

The conformal Killing spinor equation  is the form

\begin{equation}
\nabla_m\epsilon-c\Gamma_m\eta=0\, ,
\end{equation}
where $\epsilon$ is the Poincare spinor, $\eta$ the superconformal spinor and $c$ some constant depending on dimensionality. Note that, because of the product structure of the space, the spin connection appearing in the covariant derivative will have either all indices along the sphere $\mathbb{S}^a$ or all indices along the $AdS_b$. Let us denote the sphere indices by $i,j\cdots$ and the $AdS$ indices by $I,J,\cdots$.

Let us now assume that $\epsilon=\epsilon_{S}\otimes \epsilon_{AdS}$ and $\eta=\eta_S\otimes \eta_{AdS}$. We now need to write down the Dirac matrices for $\mathbb{S}^a\times AdS_b$. The decomposition depends on $a,b$ being odd or even. In order to illustrate the details, let us first consider the  \textbf{(a even, b odd)} case.  We will use some results from \cite{Lu:1998nu}. The Dirac matrices are

\begin{equation}
\Gamma_i=\gamma_i\otimes \unity\, ,\qquad \Gamma_I=\gamma_{S}\otimes \gamma_I\, ;
\end{equation}
where $\gamma_i$, $\gamma_I$ and $\gamma_{S}$ are the respectively the $\mathbb{S}^a$ Dirac matrices, the $AdS_b$ Dirac matrices and the $\mathbb{S}^a$ chirality operator satisfying $\gamma_{S}^2=\unity$. Clearly  
\begin{eqnarray}
&&\nabla_i\epsilon=\nabla_i\epsilon_S\otimes \epsilon_{AdS}\, ,\\
&&\nabla_I\epsilon=\epsilon_S\otimes \nabla_I\epsilon_{AdS}\, .
\end{eqnarray}
Then the Killing spinor equation becomes
\begin{eqnarray}
&& \nabla_i\epsilon_S\otimes \epsilon_{AdS}-c\gamma_i\eta_S\otimes \eta_{AdS}=0\, ,\\
&& \epsilon_S\otimes \nabla_I\epsilon_{AdS}-c\gamma_S\eta_S\otimes \gamma_I\eta_{AdS}=0\, .
\end{eqnarray}
This is satisfied as long as the following equations hold

\begin{equation}
\label{e1}
\nabla_i\epsilon_S=c\gamma_i\eta_S\, ,\qquad \nabla_I\epsilon_{AdS}=c\gamma_I\eta_{AdS}\, .
\end{equation}
However, note that if we $\Gamma$-trace the original equation, we find

\begin{equation}
\slashed{\nabla}\epsilon=c\,(a+b)\eta\, .
\end{equation}
Since

\begin{equation}
\slashed{\nabla}\epsilon=\slashed{\nabla}^S\epsilon_S\otimes \epsilon_{AdS}+\gamma_S\epsilon_S\otimes \slashed{\nabla}^{AdS}\epsilon_{AdS}\, ,
\end{equation}
we have that

\begin{equation}
\slashed{\nabla}^S\epsilon_S\otimes \epsilon_{AdS}+\gamma_S\epsilon_S\otimes \slashed{\nabla}^{AdS}\epsilon_{AdS}=c(a+b)\eta_S\otimes \eta_{AdS}\, .
\end{equation}
From \eqref{e1} we have $\slashed{\nabla}^S\epsilon_S=c\,a\,\eta_S$ and $\slashed{\nabla}^{AdS}\epsilon_{AdS}=c\,b\,\eta_{AdS}$. Thus

\begin{equation}
a\,\eta_S\otimes \epsilon_{AdS}+b \gamma_S\epsilon_S\otimes \eta_{AdS}=(a+b)\eta_S\otimes \eta_{AdS}\, .
\end{equation}
Hence, if we take $\epsilon_{AdS}=\eta_{AdS}$ and $\gamma_S\epsilon_S=\eta_S$, the equation is satisfied. In turn, \eqref{e1} reduces to

\begin{equation}
\nabla_i\epsilon_S=c\gamma_i\gamma_S\epsilon_S\, ,\qquad \nabla_I\epsilon_{AdS}=c\gamma_I\epsilon_{AdS}\, .
\end{equation}
These equations are precisely the Killing spinor equations on $\mathbb{S}^a$ and $AdS_b$. It should be stressed that this is true for equal radii $\mathbb{S}^a$ and $\mathbb{H}^b$, which is precisely the case when the geometry is conformal to $\mathbb{S}^{a+b}$. 

The other cases follow in a similar fashion. For instance, let us now consider   \textbf{(a even, b even)}. There are two choices for the Dirac matrices \cite{Lu:1998nu}

\begin{itemize} 

\item \textbf{i)}: $ \Gamma_i=\gamma_i\otimes \unity\, ,\qquad \Gamma_I=\gamma_{S}\otimes \gamma_I\, .
$

This is formally the same computation as the case \textbf{(a even, b odd)}, just taking into account that now $b$ is even. Thus, we have $\epsilon=\epsilon_S\otimes \epsilon_{AdS}$, where the sphere and $AdS$ Killing spinors satisfy, respectively (we set $c=\frac{1}{2}$)

\begin{equation}
\nabla_i\epsilon_S=\frac{1}{2}\gamma_i\gamma_S\epsilon_S\, ,\qquad \nabla_I\epsilon_{AdS}=\frac{1}{2}\gamma_I\epsilon_{AdS}\, .
\end{equation}
\item \textbf{ii)}: $\Gamma_i=\gamma_i\otimes \gamma_{AdS}\, ,\qquad \Gamma_I=\unity\otimes \gamma_I\, .$

This is formally the same computation as the case \textbf{(a odd, b even)},  now setting $a$ to be an even number.

\end{itemize}

The remaining cases can be easily worked out.
We thus conclude that the spaces $\mathbb{S}^a\times \mathbb{H}^b$  preserve, like $\mathbb{S}^{a+b} $, all supersymmetries, since no extra projection appears
(as long as $\mathbb{S}^a$ and  $\mathbb{H}^b$   are of equal radii),
finding that indeed it is possible to construct supersymmetric theories on $\mathbb{S}^a\times \mathbb{H}^b$. Euclidean field theories with rigid supersymmetries can be defined on $\mathbb{S}^a$ or on  $\mathbb{H}^b$  provided $a<6$ and $b<8$. Analogous restrictions will also apply to our case.\footnote{These restrictions
arise from the Nahm's classification of possible superalgebras, see \cite{Kehagias:2012fh}. For $\mathbb{S}^a$, it is possible to include the cases $a=6,7$, provided  a suitable  prescription is adopted  to circumvent some pathologies, see \cite{Minahan:2015jta} for examples.}

\end{appendix}

\end{document}